\documentclass[traditabstract]{aa} 
\usepackage{txfonts,psfig}

\begin{document}

\title{JKCS\,041: a Coma cluster progenitor at $z=1.803$ }
\titlerunning{JKCS\,041: a Coma progenitor at $z=1.803$} 
\author{S. Andreon\inst{1} \and A. B. Newman\inst{2,3} \and G. Trinchieri\inst{1}
\and A. Raichoor\inst{1,4} \and
R. S. Ellis\inst{2} \and T. Treu\inst{5}}
\authorrunning{Andreon et al.}
\institute{
$^1$ INAF--Osservatorio Astronomico di Brera, via Brera 28, 20121, Milano, Italy,
\email{stefano.andreon@brera.inaf.it} \\
$^2$ Cahill Center for Astronomy and Astrophysics, California
Institute of Technology, MS 249-17, Pasadena, CA 91125, USA \\
$^3$ Current Address: The Observatories of the Carnegie Institution
for Science, 813 Santa Barbara St., Pasadena, CA 91101, USA \\
$^4$ Current Address: GEPI, Observatoire de Paris, 77 av. Denfert Rochereau,
75014 Paris, France \\ 
$^5$ Department of Physics, University of California, Santa Barbara,
CA 93106, USA \\
}
\date{Accepted ... Received ...}
\abstract{Using deep two-color near-infrared HST imaging and unbiased grism spectroscopy,
we present a detailed study of the $z=1.803$ JKCS\,041 cluster. We confirm, for
the first time
for a high--redshift cluster, a mass of $\log M\gtrsim 14.2$ in solar units
using four different techniques based on the X-ray temperature, the X-ray luminosity,
the gas mass, 
and the cluster richness. JKCS\,041 is thus a progenitor of a local system like the Coma cluster. 
Our rich
dataset and the abundant population of 14 spectroscopically confirmed red--sequence
galaxies allows us to explore the past star formation history of this system in unprecedented detail. 
Our most interesting result is a prominent red sequence down to stellar masses as low as $\log M/M_\odot=9.8$, 
corresponding to a mass range of 2 dex. These quiescent galaxies are concentrated around
the cluster center with a core radius of 330 kpc. There are only few blue members 
and avoid the cluster center. 
In JKCS\,041 quenching was therefore largely completed by a look--back time of 10 Gyr, and we can constrain 
the epoch at which this occurred via spectroscopic age-dating of the individual galaxies. Most
galaxies were quenched about 1.1 Gyr prior to the epoch of observation. 
The less--massive quiescent galaxies are somewhat younger, corresponding
to a decrease in age of $650$ Myr per mass dex, but the scatter in age at fixed
mass is only $380$ Myr (at $\log M/M_\odot=11$).  There is no evidence for 
multiple epochs of star formation across galaxies. The size--mass relation of quiescent galaxies in JKCS\,041 is
consistent with that observed for local clusters within our uncertainties, and we place an upper
limit of $0.4$ dex on size growth at fixed stellar mass  (95\% confidence). Comparing our data on JKCS\,041 with 41 
clusters at lower redshift, we find that the form of the mass function of red sequence galaxies has
hardly evolved in the past 10 Gyr, both in terms of its faint--end slope and characteristic
mass. Despite observing JKCS\,041 soon after its quenching and the three--fold 
expected increase in mass in the next 10 Gyr, it is already remarkably similar to
present-day clusters.
}

\maketitle

\section{Introduction}

It has been recognized for many years that clusters of galaxies are valuable laboratories for
probing galaxy formation and evolution. The tightness of the red sequence of member
galaxies implies a uniformly old age for their stellar populations (e.g. Bower, Lucey \& Ellis 1992; 
Stanford et al. 1998; Kodama \& Arimoto 1997) which, together with the morphology--density
relation (e.g. Dressler 1980), leads to two important conclusions: first, present-day clusters are the 
descendants of the largest mass fluctuations in the early Universe where evolutionary
processes were accelerated; and second, cluster-specific phenomena, such as ram-pressure 
stripping or strangulation (e.g. Treu et al. 2003) played an important role in quenching
early star formation.

Even though much can be inferred from the fossil evidence provided by studies of galaxies in local
clusters, these are rapidly evolving systems and it is important to complement this information
with studies of the galaxy population at the highest redshifts. The relevant observables include
the location and width of the red sequence of quiescent members, the mass distribution
along the sequence and the various trends such as the size--mass and the age--mass
relations. Theoretical predictions of these observables are challenging because it is currently 
difficult to resolve star--forming regions in simulations that are sufficiently large to include representative 
rich clusters. Semi-analytic models do not suffer from such limitations but rely on simplifying
assumptions (e.g. on the importance of merger-induced bursts and stellar feedback) and 
tunable recipes. Even worse, because clusters are complex environments, there is not always
unambiguous insight into the relative importance of the different physical processes, which
means that it is unclear which of them should be modelled.

For these reasons, a phenomenological approach is very productive and represents
the route followed by many researchers. They have observed clusters progressively 
more distant to approach the epoch at which star formation was quenched and, particularly,
to determine the various processes involved. Quenching is thought to occur at $z\lesssim2$ 
by some semi-analytic models (e.g. Menci et al. 2008) and it is observationally at
$z\gtrsim1.2$: since at $z\simeq$1.2, the cluster population is largely
evolving passively (e.g. de Propris et al. 1999), the red sequence is already in place at the highest redshift
($z=1.4$, e.g. Stanford et al. 1998; Kodama \& Arimoto 1997), and the processes responsible
for truncating residual star formation are more effective in denser environments and for more 
massive galaxies (e.g. Raichoor \& Andreon 2012b).

Beyond z$\simeq1.4$, our knowledge of the properties of galaxy populations in clusters
becomes incomplete. 
There is a paucity of clusters with reliable masses, spectroscopically confirmed
members, and adequately deep imaging and spectroscopic data essential for determining past star
formation histories. The challenges of the complexity of astronomical
data and the associated statistical analysis compound
the paucity of the data. All this
has led to some disagreement,  for example, on the mass-dependent
distribution of the passive population (c.f. Mancone et al. 2010, 2012; Fassbender et al. 2011
vs Andreon 2013; Strazzullo et al. 2010) and whether the color-density relation reverses
from its local trends at high redshift (Tran et al. 2010 vs Quadri et al. 2012). 

We present a comprehensive analysis of the galaxies in the rich cluster JKCS\,041, which
was first
identified by Andreon et al (2009). In a previous paper in this series (Newman et al 2013,
hereafter Paper I), we derived a redshift of $z=1.803$ using HST grism spectroscopy, 
based on 19 confirmed members, a high proportion of which are quiescent. In the present
paper we derive the mass of this cluster and examine the extent of the red sequence
in the context of its likely subsequent evolution.  By considering the
color--magnitude relation, the age--mass relation and the spatial distribution of passive
members, we argue that JKCS\,041 can be regarded as the progenitor of a present--day Coma--like
cluster.

\begin{figure}
\centerline{\psfig{figure=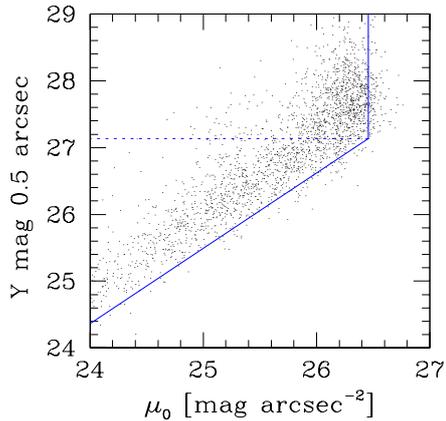,width=6truecm,clip=}}
\caption[h]{
Completeness of our $Y$--band imaging of JKCS\,041. The $Y$--band magnitude 
within a 0.5 arcsec aperture compared with the peak surface brightness. There are
galaxies with a peak surface brightness fainter than the detection threshold
($\sim 26.5$ mag arcsec$^{-2}$, vertical line), but they are undetectable with these data. Extrapolating
the trend seen at brighter magnitudes (slanted solid line), the completeness limit 
(horizontal dashed line) is 27.15 mag. Owing to uncertainties inherent in this extrapolation,
we conservatively adopt a limit that is 0.5 mag brighter, which leads to a 
limiting magnitude of $Y = 26.6$.}
\end{figure}

The plan of the paper is as follow: in \S2 we briefly review the HST data; details
are available in Paper I. In \S3 we estimate the mass of JKCS\,041 using various
diagnostics, including its X-ray properties and optical richness and compare
these with those of other high--redshift clusters. In \S4 we analyze 
the ages and colors of the passive members in detail, demonstrating an important
result: the red sequence is fully populated down to low stellar
masses. We also examine the age--mass relation and the spatial distribution
of red members in light of conflicting results presented for other clusters.
In \S5 we discuss our results on JKCS\,041 in the context of a sample of other
clusters at lower redshift and consider the most likely evolutionary picture. We
discuss the implications in \S6.

Throughout this paper, we assume $\Omega_M=0.3$, $\Omega_\Lambda=0.7$, 
and $H_0=70$ km s$^{-1}$ Mpc$^{-1}$. Magnitudes are in the AB system.
We use the 2003 version of Bruzual \& Charlot (2003, BC03 hereafter) stellar population synthesis
models with solar metallicity and a Salpeter initial mass function (IMF).
We define stellar masses as the integral of the star formation
rate, and they therefore include the mass of gas processed by stars and returned to the interstellar medium.
In \S4.4 and \S4.5, for consistency with Paper~I, we instead use stellar masses that count only the
mass in stars and their remnants; these masses are systematically lower by 0.12~dex.

\section{Data}

The primary data used in this paper were derived from the HST imaging and spectroscopic 
campaign (GO: 12927, PI: Newman) discussed in Paper I. Only a brief summary is provided,
and we refer to Paper I for more details.

The imaging data consist of photometry in two bands, F105W (for a total observing time of 2.7 ks) 
and F160W (4.5 ks), denoted Y and H, respectively, derived using the SExtractor code
(Bertin \& Arnouts 1996). Total galaxy magnitudes refer to isophotal-corrected magnitudes, while colors
are based on a fixed 0.5 arcsec diameter aperture, which is well suited for faint galaxies at the 
cluster redshift. The photometric data for red--sequence galaxies in the cluster is
complete to $Y$= 26.6 mag (see Figure 1). As a control sample 
(to estimate the number of back/foregound galaxies), we selected an 
area of 29.78 arcmin$^2$ in the GOODS South area observed in a similar manner as
part of the CANDELS program (GO: 12444/5, PI: Ferguson, Riess \& Faber,  
Grogin et al. 2011; Koekemoer et al. 2011; we used the images distributed in
Guo et al. 2013). 

The spectroscopic data were taken with the WFC3/IR G102 and G141 grisms
at three epochs (and hence orientations) to reduce contamination. The spectra
sample from 3000 to 6000 \AA \ in the rest-frame of galaxies at $z=1.8$.
Redshifts were published in Paper I for 63 emission--line galaxies brighter than $H=25.5$ mag 
and 35 absorption--lines galaxies brighter than $H=23.3$ mag. Nineteen of these
are cluster members. The absorption--line galaxy sample of Paper I has a
pre-selection based on photometric redshift estimates, such that only galaxies 
with $1.4<z_{phot}<3$ were considered; no such cut was applied for emission--line galaxies.

For this paper and for the purpose of checking completeness, we also extracted and fit
the spectra of those few bright ($H<22.5$ mag) galaxies irrespective of their photo-z
within $1.2^2$ times the area enclosed by the outermost detected X-ray isophote (1.8 arcmin$^2$, 
about
$0.5$ Mpc$^2$). These targets were not considered in Paper I. The
successfully extracted spectra indicate redshifts $z<1.2$, in agreement with the 
photo-z pre-selection. Of the four unsuccessfully extracted spectra, two represent
galaxies that are very likely too extended to be at high redshift. The two remaining galaxies
have a low photo-z and are probably foreground galaxies. Therefore, the sample of galaxies 
with $H<22.5$ mag with spectroscopic redshifts represents a complete sample
of cluster members.

In this paper we complement the HST data with data from the WIRCam Deep Survey (WIRDS) survey (Bielby et al.
2012), which enables us to estimate the mass of JKCS\,041 from its richness (see \S 3).
WIRDS provides deep $z'$ and $J$ photometry of a 1 deg$^2$  field around JKCS\,041 to
a limit of $z'=25$ mag. Our earlier papers characterized the red-sequence relation for JKCS\,041 on the basis
of these data (Andreon 2011, Raichoor \& Andreon 2012a). 

\section{Mass of the JKCS\,041 cluster}

We now turn to the mass of JKCS\,041  which can be estimated in four ways.

\begin{enumerate}
\item
Using the X-ray temperature--mass relation under the assumption that the locally determined 
relation evolves self--similarly. Although this assumption has been challenged by, for
example, 
Andreon, Trinchieri, Pizzolato (2011 and references therein), the implications are not
yet precisely quantified given that samples with a known selection function would be required 
(Pacaud et al 2007;  Andreon, Trinchieri \& Pizzolato 2011; Maughan et al. 2012). 

\item
Using the $L_X$-mass relation (Vikhlinin et al. 2009), again 
assuming a self--similar--inspired evolution.

\item
Using the gas mass, assuming that the gas fraction has not changed in the last 10 Gyr.

\item
Using a local richness-mass relation (e.g. Johnston et al. 2007; Andreon \& Hurn 2010). 
The mass estimated from the cluster richness is probably relatively independent of evolutionary effects, 
given that the color evolution of red members is reasonably well understood (e.g. De Propris et al. 1999, see also \S 5.1). 
\end{enumerate}

For the first three methods we used the results from the Chandra observations reported by Andreon et
al. (2009), rescaled to the current redshift and to the same aperture $r_{200}$ assuming a Navarro, Frenk
\& White (1997) profile of concentration $3-5$.  We find $\log M_{200}/M_\odot=14.6\pm0.5$ from the
X-ray temperature.  Using the X-ray luminosity in the [0.5-2] keV band within $r_{500}$ 
and the Vikhlinin et al.
(2009) calibration, we find $\log M_{200}/M_\odot\approx14.45$. For this latter estimate, there is an
uncertainty of $\ge 0.2$ dex, mostly arising from the scatter in $L_X$ at a given mass (Vikhlinin et al.
2009, $\ge$ 0.11 dex) with $0.08$ dex arising from uncertainties in the count rate and temperature.
Assuming a 10 \% gas fraction, typical of nearby clusters (Vikhlinin et al. 2009), we find
$\log M_{200}/M_\odot\approx14.4$ from the gas mass within $r_{500}$ (Ettori et al. 2004). 
This gas--based mass has an error of $\ge 0.10$ dex, derived from 
a $0.08$ dex arising from the uncertainties in the gas--mass profile and the local Universe $\sim 0.06$ dex
scatter of the gas fraction (Andreon 2010), to which we need to add an unknown uncertainty because 
of the evolution of the gas fraction. 
Adopting the Universe baryon fraction as value of the gas fraction, 
we find $\log M_{200}/M_\odot\approx14.2$.
Given that it is unlikely that the gas fraction exceeds
the Universe baryon fraction, at face value $\log M_{200}/M_\odot\approx14.2$
is a reasonable lower limit of the JKCS\,041 mass. 
Finally, from the X-ray profile and spectrum normalization, we computed
a gas mass within 30 arcsec of $3.9
\ 10^{12}$ $M_\odot$, fully consistent with the upper limit derived from SZ observations in Culverhouse
et al. (2010) and extrapolation of lower redshift scaling relations.

The fourth method uses the large area sampled by the WIRDS images and, using the calibration of Andreon \& Hurn (2010), 
is based on the number of red-sequence galaxies brighter than $M^{z=0}_V=-20$ mag within $r_{200}$, 
correcting for line-of-sight contamination using galaxies at projected radii $6.5<r<9.1$ Mpc. 
To derive the equivalent local luminosity $M^{z=0}_V$, we assumed passive evolution using the BC03 models
with a simple stellar population formed at redshift $z_f=3$, a Salpeter IMF and solar metallicity, although the results
are not sensitive to these parameters. With 20 red--sequence galaxies brighter than $K_s=21.99$ mag within 1.43 Mpc, 
JKCS\,041 is within the lower half of the mass range $13.7<\log M/M_\odot< 15$ discussed by Andreon \& Hurn (2010).
Using their eqn. 18, we estimate $r_{200}$=0.76 Mpc, which contains 17 members, and derive 
$\log M_{200}/M_\odot=14.25\pm0.29$. This uncertainty takes into account the intrinsic scatter in mass
at a given richness (the dominating term), the limited number of members, and the uncertainty of
the calibrating relation.

As can be seen, the richness--based mass estimate agrees reasonably well 
with the mass estimates:
$\log M_{200}/M_\odot=14.6\pm0.5$ from the X-ray temperature, $\log M_{200}/M_\odot\approx14.45$ from the X-ray
luminosity, and $\log M_{200}/M_\odot\approx14.4$ from the gas mass.  
This is a unique achievement for such a high--redshift cluster. All four determinations agree to
better than 1 combined standard deviation, indicating that JKCS\,041 is an intermediate--mass cluster by the
standards of local systems, and a massive cluster at such high redshift. In fact, a typical $z=1.8$ cluster with
$\log M_{200}/M_\odot\approx 14.25$ will
grow in mass (Fakouri et al.  2010) to become a $\log M_{200}/M_\odot\approx 14.75$ cluster locally, albeit
with an uncertainty due to the stochastic nature of the mass growth in a hierarchical Universe. Nonetheless, our
mass estimates suggest that  JKCS\,041 is a credible ancestor of a present-day, Coma-like cluster. 

\subsection{Comparing JKCS\,041 with other high--redshift clusters}

In light of these results, we now compare the mass of the JKCS\,041 cluster
with that of other high--redshift systems in the literature.

Following the same method and local calibration as used for
JKCS\,041, Stanford et al. (2012) derived a  
mass of $\log M_{200}/M_\odot\approx 14.5\pm0.1$  from $L_X$ for 
IDCS J1426.5+3508 ($z$=1.75), similar to the value derived for JKCS\,041,
but with a smaller uncertainty, because the authors 
determined it on statistical measurement errors alone.
Given that both clusters are at similar redshift, a comparison of the 
X-ray luminosities obtained in comparable apertures is illustrative. As neither the 
X-ray temperature nor the X-ray surface brightness radial profile is measurable for 
IDCS J1426.5+3508, we can only compare the X-ray count rates in similar apertures 
($1$ arcmin for both clusters). We found a factor two in count rate,
which, divided by the slope of the $L_X$-mass relation ($1.6$), gives
a factor 1.5 in mass, with IDCS J1426.5+3508 being more massive. However, allowing for
the scatter in $L_X$ at a given mass, the masses of the two clusters are statistically 
indistinguishable.

To compare the mass estimate based on richness, we also need to compare the relative 
fraction of red galaxies. Since both clusters have similar richness within
1 arcmin (Andreon 2013), their masses are also expected to be similar. 
We note, however, that while the bright galaxies in JKCS\,041 are all red (Raichoor \& Andreon
2012, Andreon 2013 and sec 4.1), there are several blue galaxies in IDCS J1426.5+3508 (see Andreon 2013),
indicating a lower richness $n_{200}$. Inspection of the color-magnitude relation of the two
clusters reinforces this impression: while all bright galaxies are on the
red sequence in JKCS\,041, the red sequence of IDCS J1426.5+3508 seems poorly populated,
and even more so if some allowance is made for background/foreground contamination
(the almost totality of red sequence IDCS J1426.5+3508 galaxies has an unknown membership).

By contrast, the CL J1449+0856 structure at $z=1.99$ (Gobat et al. 2009) has a much lower mass.
Only the brightest cluster member is more luminous than $M^{z=0}_V=-20$ mag.
Because of the paucity of luminous members for CL J1449+0856, there are no 
low-redshift structures in the  
sample of Andreon \& Hurn (2010), which we may use as a calibrator, except for possibly 
the NGC4325 group, with two galaxies brighter than $M^{z=0}_V=-20$ mag and mass 
$\log M_{200}/M_\odot = 13.3\pm0.3$. This value is lower than, although formally consistent with, 
that estimated by Gobat et al. (2011) using $L_X$ and the estimate from Strazzullo et al. (2013)
derived from the total stellar mass within $r_{500}$ (using the local calibration of Andreon 2012).  
Both estimates indicate a mass an order of magnitude lower that 
JKCS\,041 and IDCS J1426.5+3508, consistent with its smaller core radius (20 kpc, 
Strazzullo et  al. 2013).

\section{The galaxy population of JKCS\,041}

\begin{figure}
\centerline{\psfig{figure=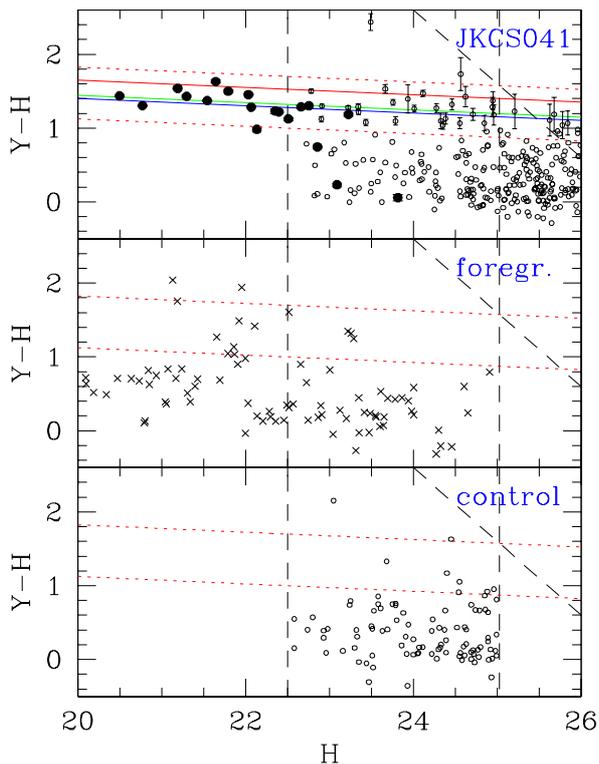,width=8truecm,clip=}}
\caption[h]{Color-magnitude diagram. {\it Upper panel:} 
JKCS\,041 spectroscopic members (solid dots) and all
$22.5<H$ galaxies within the JKCS\,041 area (open dots). 
The locus of an SSP with $z_f=5, 3, 2.5$ is shown for comparison (solid lines,
from top to bottom). For clarity, error bars are only shown for $Y-H>1$ mag galaxies.
{\it Central panel:} spectroscopic non-members. {\it Bottom panel:} 
control field galaxies with $22.5<H<25$ mag. In all panels the JKCS\,041 
$Y$-band completeness limit is indicated with a slanted dashed line ($Y=26.6$ mag),
along with the corresponding $H$ completeness limit (vertical dashed line, $H=25$ mag) for
the reddest red-sequence galaxies. The JKCS\,041 spectroscopic completeness limit of $H=22.5$
and the locus of red--sequence galaxies (red dashed corridor) are also shown.
}
\end{figure}

\begin{figure}
\centerline{\psfig{figure=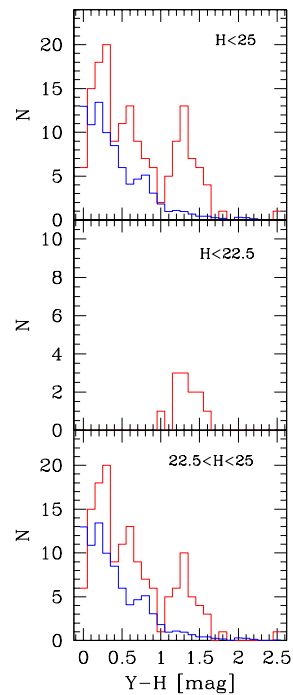,width=4truecm,clip=}}
\caption[h]{Color distribution in the cluster (red) and control
field (blue) for all galaxies (top panel), bright galaxies (middle panel) and
faint galaxies (bottom panel). Since membership is available on an individual
basis for the bright galaxies, no control field is required for the middle panel.
Colors are corrected for the slope of the color--magnitude
relation assuming a slope of 0.05, consistent with the JKCS\,041 data and with 
slopes observed at $z=0$.}
\end{figure}

\subsection{A fully--populated red sequence}

In this section we isolate the red--sequence members of JKCS\,041 and show,
for the first time in a high-redshift cluster, the presence of quenched galaxies over
a 2 dex range in stellar mass at a look--back time corresponding to 
only 3.5~Gyr after the Big Bang.

As described earlier, our catalog is spectroscopically complete for galaxies with $H<22.5$
mag that lie within the outermost detected X-ray isophote enlarged by a factor of 
1.2\footnote{This region (i.e., the factor $1.2\times$) was chosen as a compromise
between maximizing the number of enclosed red-sequence cluster members and minimizing
the number of random interlopers, as judged from the control field.}. For these bright galaxies,
we have proof of membership for each galaxy. For fainter galaxies with $22.5 < H < 25$ mag, 
however, we must statistically estimate the contamination by galaxies unrelated to the cluster using
the GOODS-S control field.

The top panel of Figure 2 shows all galaxies within this area that are either 
confirmed spectroscopic members (15, solid points) or whose membership is unknown 
(open circles).  
Additionally, we plot the four remaining spectroscopic members found outside this
area (two are red and two are blue). The figure demonstrates that at $H<22.5$ mag, all members
are red galaxies, in agreement with previous results based on a purely
statistical analysis of membership (Andreon \& Huertas-Company 2011; Raichoor \& Andreon 2012a).
We detected in our Chandra data one red-sequence member
($L_X\approx 6 \ 10^{42}$ erg s$^{-1}$ [0.5,2] keV) that might accomodate a low
luminosity AGN. No other members are detected, and for them we derive
$L_X<  \ 10^{43}$ erg s$^{-1}$. 
None of the red-sequence galaxies, nor their stack were detected in the SWIRE
data (Paper I).
The red sequence is much more prominent than other high--redshift clusters, 
in particular the $z=1.75$ cluster IDCS J1426.5+3508 (Stanford et al. 2012).
The $Y-H$ color of the JKCS\,041 red sequence is consistent with expectations
for a simple stellar population (SSP) formed at $z_f=2.5 - 3.0$, based on BC03 models
with a Salpeter IMF and solar 
metallicity\footnote{Stellar ages derived from the grism spectra are discussed in \S4.4.}.  
Most of the spectroscopic non-members, shown in the middle panel, 
are bluer than the cluster red sequence. Many of the red non-members are also at high redshift, but 
do not meet the membership criterion defined in Paper~I ($|z-z_{clus}|<0.022$). 

The bottom panel of Figure~2 compares the color distribution in the GOODS-S control field. Although
this field covers 29.78 arcmin$^2$, for clarity we only show galaxies drawn from a solid angle
equal to the cluster field. The comparison reveals more galaxies with $Y-H\sim 1.3$ mag 
in the cluster, indicating that the cluster red sequence is very well defined at all luminosities: 
at the bright end, where we were able to eliminate non-members using grism spectroscopy, 
and at the fainter end, where it is evident to $H=25$ mag. This is the first time that the red sequence has
been traced in clusters to such luminosities at high redshift, which is due to the depth of our $Y$-band 
observations.

More quantitatively, Figure~3 shows the color distribution in the cluster and control fields 
(red and blue lines, respectively) for the entire sample ($H<25$ mag) and for two subsamples 
divided at $H=22.5$ mag. All colors are corrected for the color-magnitude slope. Since
spectroscopic membership is known for bright galaxies,  background contamination is not
an issue for $H<22.5$ mag. The data reveal a clear excess of galaxies at $Y-H\sim1-1.7$ mag in both
the brighter and fainter subsamples. This excess (28 galaxies) far exceeds 
the observed scatter 
in our control field across regions of
JKCS\,041 solid angle and expectations from
cosmic variance (Moster et al. 2011).  The slope-corrected 
$1<Y-H<1.7$ mag range is our working definition of red-sequence galaxies. With this definition
one spectroscopic member (ID447) brighter than $H=22.5$ mag is blue (by a small margin, 0.02 mag), 
as are three fainter galaxies (693, 531, 332). These latter systems were classified as star-forming in 
Paper I on the basis of their UVJ colors, and the two bluest show emission lines in their spectra.

Figure~3 also shows an excess of bluer galaxies ($Y-H < 1$ mag) galaxies in the direction of JKCS\,041.
Some of these are most likely members of JKCS\,041 (including the three spectroscopically-confirmed blue members),
while the others belong to structures along the line of sight. The area around JKCS\,041 is 
known to contain several foreground groups and structures (Le Fevre et al. 2005; Andreon et al. 
2009; Andreon \& Huertas-Company 2011 and Paper I). These excess blue galaxies 
are therefore expected and are not relevant for the remainder of this paper, which is
focused on the red cluster members.

In summary,  the red sequence of JKCS\,041 is populated over an unprecedented 5 magnitudes in
luminosity, indicating that  quenching was already effective 10 Gyr ago. Because semi-analytic
models (e.g. Menci et al. 2008) predict a depopulated red sequence at bright magnitudes,
our observations indicate that the epoch of quenching may be earlier than these models assume. 
Similarly, McGee et al. (2009)  predicted that environmental quenching is negligible by $z\sim1.5$, while 
our comparison between JKCS\,041 and the $z\sim1.8$ field population clearly shows
that environmental processes are responsible for quenching more than half of the
cluster members (Paper I, Figure 8). 

\begin{figure}
\centerline{\psfig{figure=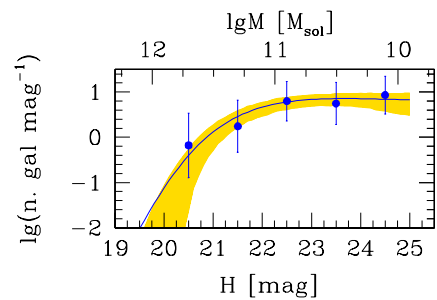,width=6truecm,clip=}}
\caption[h]{Luminosity (bottom axis) and stellar mass (top axis) 
function of red--sequence cluster members. The solid line
is the mean model fitted to the individual galaxy data, 
while the shading indicates the 68\% uncertainty
(highest posterior density interval). Points and approximated
error bars are derived 
by binning galaxies in magnitude bins, 
adopting approximated Poisson   
errors summed in quadrature, as commonly done in the literature.
The top axis indicates the corresponding stellar mass,
based on our standard BC03 SSP model with $z_f=3$.
}
\end{figure}

\subsection{The mass function of JKCS\,041}

The stellar mass function (MF) represents the zeroeth--order statistic of a galaxy
sample, giving the relative number of member galaxies as a
function of their mass. 

Figure~4 shows the mass function of red--sequence galaxies in JKCS\,041 with
$\log M/M_\odot >9.8$. (As in \S 4.1, we considered galaxies located within the outermost
X-ray isophote enlarged by a factor of 1.2). Here we converted the observed  
$H$-band luminosity to stellar mass assuming our standard BC03 SSP model with $z_f=3$. 
After accounting for the different mass definitions (\S1), masses derived from this procedure 
are consistent with those derived from fits to the grism spectroscopy and 12-band photometry (Paper I):
the median difference and interquartile range are $0.02$ dex and $0.10$ dex, respectively,
for red--sequence galaxies.

The luminosity function (LF) was derived by fitting a Schechter (1976)
function (for the cluster galaxies) by statistically subtracting the field population for $22.5<H<25$,
whose log abundance was taken to be linear.
The relevant likelihood expression was taken from Andreon, Punzi \& Grado (2005), 
which is an extension of the Sandage, Tammann \& Yahil (1979) likelihood  
when a background component is present. Uniform priors were taken for the five parameters. 
For display purposes, we also computed approximate LFs, by binning galaxies in magnitude  
and adopting approximated errors as usual (e.g. Zwicky 1957, Oemler
1974). These approximate LFs are
plotted in the figures as points with error bars, although the fit itself is 
based on the unbinned data.

For the JKCS\,041 red sequence, we derive a characteristic magnitude $H^*=21.6\pm0.5$ mag,
a characteristic mass $\log M/M_\odot \sim 11.2$,
a faint--end slope $\alpha=-0.8\pm0.2$ and a richness $\phi^*=18\pm8$ galaxies mag$^{-1}$.
The luminous end is constrained by the lack of $H<20$ mag galaxies and the paucity
of $H\sim21.5$ mag galaxies. The constraints on the slope arise both from
the nearly uniform distribution over $H=22-25$ mag and
our knowledge of the individual spectroscopic membership of the brighter galaxies,
which strongly reduces the degeneracy between faint--end slope and characteristic magnitude.
The data are complete down to $\log M/M_\odot \sim 9.8$. 

\begin{figure}
\centerline{\psfig{figure=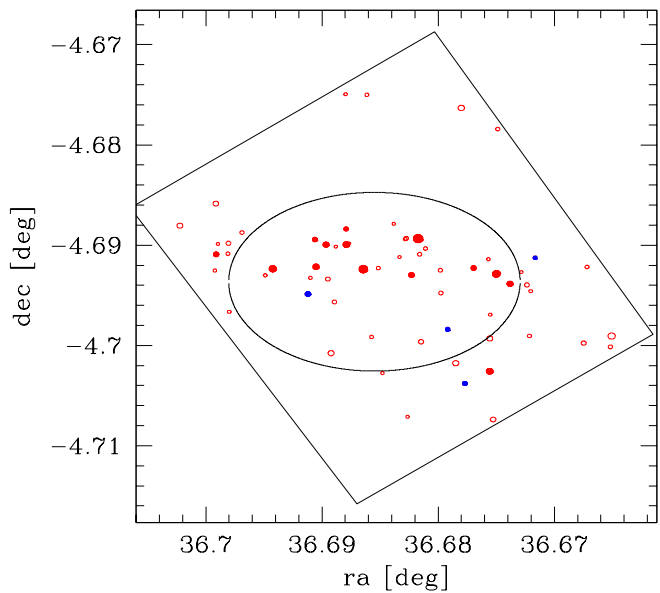,width=9truecm,clip=}}
\caption[h]{Spatial distribution of red--sequence galaxies (red
points), with symbols encoding brightness (brighter galaxies are 
indicated by larger points) and membership (spectroscopic members are indicated 
by filled points). The four blue spectroscopic members are also
shown as blue points. The ellipse indicates
the outermost detected X-ray isophote, and the slanted rectangle encloses the 
full--depth WFC3 field of view.
}
\end{figure}

\subsection{Nature of the color--density relation at high redshift}

In this section we show that the quenched galaxies in JKCS\,041 are located
preferentially at the center of the cluster potential, as is the case in low--redshift clusters.
Figure~5 shows the spatial distribution of the red--sequence member galaxies (red
points) compared with the extent of the X-ray emission (ellipse). 
Red--sequence galaxies (and the marginally
blue galaxy ID447) are clearly concentrated within the zone of detected 
X--ray emission. The three blue members, in contrast, are located away from the cluster center.
Clearly, the physical mechanism that quenched star formation 
in JKCS\,041 left only a handful of star-forming galaxies that avoid the center
as in nearby clusters
(e.g. see Figure~5 in Butcher \& Oemler 1984 and, for the Coma cluster, 
Figure~4 in Andreon 1996).
  
\begin{figure}
\centerline{\psfig{figure=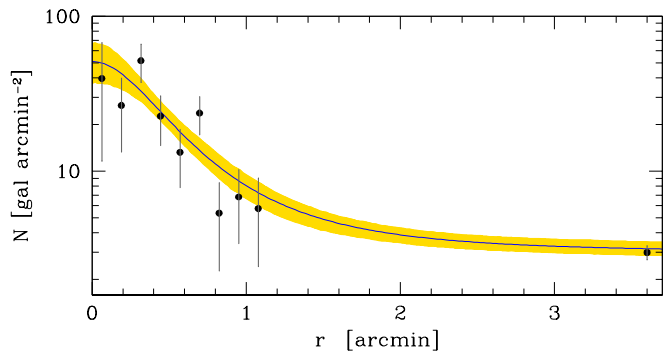,width=9truecm,clip=}}
\caption[h]{Radial profile of red--sequence galaxies. The solid line
is the mean fit and the shading indicates the 68\% uncertainty. 
The control field data point has been arbitrary put at $r=3.6$ for 
clarity. 
}
\end{figure}

Figure~6 shows the circularized radial ($=\sqrt{ab}$) distribution of the red--sequence galaxies. 
Galaxies were counted within elliptical annuli and fitted with a beta model 
(Cavaliere \& Fusco-Femiano 1976), accounting for Poisson fluctuations and for background
contamination estimated using the GOODS-S control field. We also accounted for
the boundary of the WFC3 field of view and took uniform priors on the model parameters.
The shaded region in Figure~6 shows the 68\% error (highest posterior density
interval). As in \S4.2, the binned data are shown only for display purposes.

We found a core radius $r_c=0.66\pm0.18$ arcmin (330 kpc), 
a radial slope $\beta=1.6\pm0.6$,  and a central density value of $48\pm13$ galaxies arcmin$^{-2}$.
This is similar to the core radius $r_c=0.62\pm0.13$ arcmin derived from the X-ray 
observations (Andreon et al. 2009),  which was derived with 
$\beta$ fixed at $2/3$. The core radius of JKCS\,041  
resembles that in present-day clusters, which typically have $r_c \approx 250$ kpc
 (e.g. Bahcall 1975). The absence of star-forming galaxies in the core agrees with 
Raichoor \& Andreon (2012a,b) who showed, based on a statistical analysis (i.e. 
without knowledge of individual galaxy membership), that the present-day star 
formation-density relation is already in place in JKCS\,041 and that the processes 
responsible for the cessation of star formation in  clusters are effective already at
high redshift\footnote{As mentioned in these papers, the analysis assumed 
$z_{phot}=1.8$ for consistency with galaxy colors, and thus requires no revision, 
even though the JKCS\,041 points are plotted at $z_{phot}=2.2$ for consistency
with earlier estimates of the cluster redshift.}.

A key result is that in JKCS\,041 there is no evidence for an inversion of the local 
color--density relation, as has been suggested in the densest regions 
at high-$z$ (Elbaz et al. 2007) and in a candidate cluster at similar redshift
(Tran et al. 2010; but see Quadri et al. 2012, who instead claimed
an elevated fraction of quiescent objects in the center of the same cluster).
Nor do we find that 40\% of galaxies in the cluster core 
have an intense episode of star formation (Brodwin et al. 2013).

\begin{figure}
\centerline{\psfig{figure=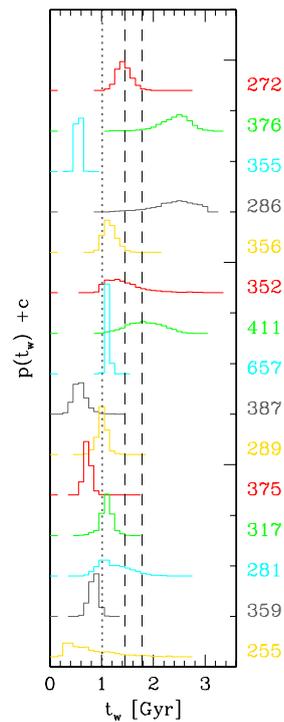,width=4truecm,clip=}}
\caption[h]{Weighted age distribution of the spectroscopic members of JKCS\,041that lie on
the red sequence, sorted in order of decreasing mass (more massive toward the top). The three vertical
lines correspond to $z_f=2.5, 3, 3.5$ Gyr from left to right. }
\end{figure}

\subsection{Age--mass relation}

In this section we measure the mean age--mass relation and its scatter for the red--sequence 
cluster members. The data quality and the proximity to the last major episode of star formation
allow us to constrain for the first time the distribution of ages at a fixed mass. In particular, we can
test for distributions that would indicate multiple epochs of star formation, across galaxies.

In Paper I, we fit grism spectra covering the rest-frame wavelength
range $\sim 3000$ to $\sim 6000$ \AA, along with photometry from $u$ to [4.5] filters,
with BC03 $\tau$ models characterized by a star formation time onset $t_0$, the $e$-folding time $\tau$, 
and dust attenuation. The galaxy spectra and spectral energy distributions (SED)
basically constrain some degenerate combination of $t_0$ and $\tau$.  Furthermore, the 
average stellar age of a stellar population is in general younger than $t_0$.
Therefore, following Longhetti et al. (2005), in the present paper we refer
to ages $t_w$ that are weighted by the star-formation history (SFH) and thus
represent the average luminosity-weighted age of the population:

\begin{equation}
t_w = \frac{\int t SFH}{\int SFH}.
\end{equation}
Eq. 1 has the following analytic solution for exponentially declining
$\tau$ models of the SFH:
\begin{equation}
t_w = \frac{t_0 }{1-e^{-t_0/\tau}} - \tau.
\end{equation}

Because $t_0$ is significantly higher than $\tau$ for the red--sequence galaxies,
$t_w$ typically differs from $t_0$ by $\approx 10\pm5$\%, with a maximal difference of 30\%.

Figure~7 shows the probability distributions for the SFH-weighted age of the red-sequence galaxies 
with spectra (technical details can be found in Appendix A) . As mentioned before, the
latter is essentially a mass-complete sample at $H < 22.5$ mag (2.5 mag brighter
than the limit considered in earlier sections, because of the requirement of a spectrum)
and a statistically representative sample to a limit that is 0.8 mag fainter (i.e. 0.3
dex lower). Galaxies in Figure~7 are ordered by mass, with more
massive objects at the top. Inspection of the figure qualitatively suggests that 
lower-mass galaxies have a younger age, about 1.0 Gyr,
whereas massive galaxies have a broader age range and are possibly 
older. A notable exception is galaxy ID355, which is significantly younger than other galaxies 
of similar mass; as already remarked in Paper I, the presence of manifestly stronger Balmer lines
in the grism spectrum and a brighter rest-frame UV continuum indicate that the age difference is genuine.

\begin{figure}
\centerline{\psfig{figure=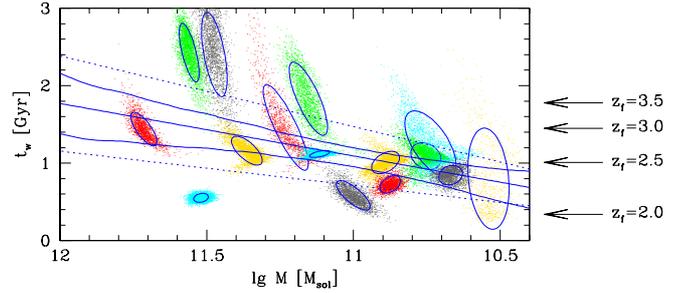,width=9truecm,clip=}}
\caption[h]{Age-mass relation for red--sequence galaxies in JKCS\,041.
The clouds of colored points show the probability distribution for each individual galaxy, following
the color coding in Figure~7; ellipses show Laplace approximations.
The straight solid line is the mean age-mass relation derived from these
posteriors, based on the galaxy spectra and photometry. 
Fan-shaped lines delineate the error on
the mean model (68 \% highest posterior interval), while
the dotted lines indicate the intrinsic scatter ($\pm1\sigma$).
}
\end{figure}

To investigate how the mean age depends on mass and to quantify
the scatter in age at a given mass, we fit a linear relation between $t_w$ and mass.
Our fitting procedure accounts for the error covariance between the mass and age
measurements, for the presence of an intrinsic (possibly non-Gaussian) scatter at a given mass,
and for the possible presence of outliers
from the mean relation (such as ID355, for example; for details see Appendix A). 

Figure~8 shows the data (colored clouds of dots), the mean fitted model (central solid
line), its 68\% error (solid, fan-shaped lines), and the intrinsic scatter ($\pm 1\sigma$, dotted lines).
We verified that our fitting procedure correctly recovers the underlying parameters based on tests
with simulated data, and we emphasize the importance of such tests in situations where the
measurement errors are correlated and intrinsic scatter is significant.

The mean age of the red--sequence galaxies at $M_* = 10^{11}$ $M_\odot$ is well determined:
$1.1\pm0.1$ Gyr, i.e., $2.5<z_f<2.7$. Note that this small uncertainty includes the observational
uncertainties (the finite number of galaxies and their scatter, errors on photometry and 
spectroscopy) but only a few model uncertainties: we marginalized
over dust attenuation but assumed a fixed (solar) metallicity, a simple
(exponential) shape for the SFH, and the validity of the BC03 model spectra.

The slope of the age--mass relation is $0.67\pm0.35$ Gyr per mass dex, consistent with that seen in Coma 
(e.g. Nelan et al. 2005). Because this slope is positive,
it demonstrates that the mean stellar age depends on galaxy mass, with 95\% probability.
Red--sequence galaxies with $\log M/M_\odot\sim11.5$ are 1.4 Gyr old (i.e., $z_f=2.9$ if they were simple
stellar populations, SSPs), while those with $\log M/M_\odot\sim10.5$ are 0.7 Gyr old ($z_f=2.3$). This
age difference agrees with results found in Paper I (sec 5) by stacking the continuum-normalized spectra
of the quiescent cluster members. Compared with the  analysis of stacked
spectra in Paper~I, the present model
allows for a more general SFH (an exponentially declining model, rather than a SSP) and a spread
in ages at a given mass,
but assumes a fixed solar metallicity. The analysis in Paper~I showed that the spectral
trends are too strong to be interpreted as arising from metallicity differences alone, justifying
our current choice.  We note that the mean age measured here agrees with that inferred from the location
of the red sequence in the color--magnitude
plane (Figure~2), although this is not an independent test since the $Y-H$ color 
is used in both age estimates.

There is a statistically significant intrinsic scatter $\sigma_{t_w|M}$ at a given mass, 
that is beyond that
expected from the measurement errors. The inferred age scatter is $38\pm9$\%,
i.e., $370\pm80$ Myr at the median mass of our sample, $\log M/M_\odot\sim11$. Note
that this small scatter cannot be due simply to a bias in our spectroscopic sample
because of the high completeness discussed in \S 2. This age scatter is larger  
than the $160\pm30$ Myr previously estimated by Andreon (2011) at the 2.5 (combined) sigma level. 
This is mostly due to the high photometric redshift $z_{phot}=2.2$ assumed in that work, which
implied that the red--sequence color $z'-J$ probes a bluer rest-frame band than it does at the true
redshift $z_{grism}=1.803$. This is significant because bluer bands are more sensitive to age
differences.

In JKCS\,041 we can probe closer to the epoch of star formation than in earlier
studies of lower-redshift clusters, and yet we find a rather synchronous star formation history, 
with a spread in ages of only $370$ Myr. In particular, the precise spectroscopic ages 
(typical errors of $120$ Myr, i.e., one-third of the age spread) allow us to resolve the distribution
of ages and to verify whether there were multiple epochs of activity across galaxies. 
To test for this, we first 
computed standardized residuals around the fitted model:
\begin{equation}
stdz \ resid_i = \frac{O_i-E_i}{\sqrt{O^2_{err,i}+\sigma^2_{intr} \ E^2_i}} \ ,
\end{equation}
where $O_i$ and $E_i$ are the observed and expected value of the $i^{th}$ galaxy
and $O_{err,i}$ and $\sigma_{intr}$ are the age error and the intrinsic
scatter around the mean relation. 
We then quantified possible deviations using a 
quantile-quantile plot, using the Weibull (1939) estimate of the quantile. 
This approach has the advantage that it does not bin the data and 
is a powerful tool to check for non-Gaussian distributions of an unknown nature
(as opposed to looking for a specific type of non-Gaussianity, such as
asymmetry). Similar to the Kolmogorov-Smirnov test, this test basically compares  
the quantile of the observed values with those of a Gaussian distribution. 
We repeated this procedure 
for 5000 simulated samples each composed of 14 galaxies
drawn from the model fitted on the real data, 
and with uncertainties and covariance
as real data.
75\% of random samples drawn from the null (Gaussian) model show
larger discrepancies than those seen in the data. (This Bayesian $p$-value is computed by
Markov Chain Monte Carlo, which allows us to account for the uncertainty in 
the relation scatter, intercept, and slope,
see Andreon 2012a,b for details). This indicates
that the age distribution (across galaxies) 
at fixed mass is indeed consistent with a Gaussian distribution,
lending observational support to the widespread assumption that red--sequence cluster 
members follow a synchronous star formation history.

\subsection{Galaxy size--mass relation}

\begin{figure}
\centerline{\psfig{figure= 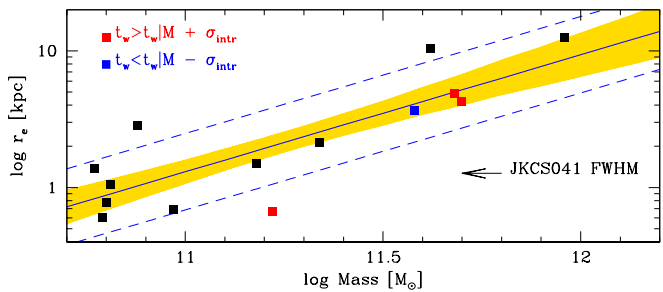,width=9truecm,clip=}}
\caption[h]{Size--mass relation for JKCS\,041. Lines show the mean fitted
model and its uncertainty (solid line and yellow band), along with the intrinsic scatter (dotted line).
Points are color-coded according to their location in the age--mass plane (Fig.~7):
older (red) or younger (blue) than the mean relation by at least $1\sigma_{intr}$, with
black points indicating galaxies within $\pm1\sigma_{intr}$ the mean age--mass relation.
}
\end{figure}

The evolution of  the size--mass relation has been a topic of great interest over
the past several years (see Paper I and references therein). Here we investigate whether
the sizes of the red--sequence members of JKCS\,041 at a given mass depend on their age. Such
a scenario might be expected if the size--mass relation evolves because of the progressive quenching
of larger galaxies over time.

Figure~9 shows the size--mass relation for the red--sequence 
galaxies in JKCS\,041. We have excluded ID387, which has a bulge plus
disk morphology and is not quiescent according to the $UVJ$ criterion used in Paper~I
(in other words, its red color is attributed to dust rather than age).
Circularized effective radii were derived in Paper I
by fitting the WFC3 $H$-band surface brightness profiles with a Sersic model. The stellar mass was
likewise derived in Paper I from spectra and photometry.
For several galaxies the derived radii are smaller than the HST resolution and the WFC3 sampling; for these
very compact systems, the assumption of a Sersic profile within
the effective radius may be particularly important.

We fit the size--mass relation of JKCS\,041 using a linear model with
intrinsic scatter, deriving
\begin{equation}
\log r_e = 0.11\pm0.09 \ +(0.86\pm0.12)(\log M -11)
\end{equation}
(solid line in Fig.~9)
and an intrinsic scatter of $0.27\pm0.06$ dex in log $r_e$ at fixed mass.
Here we adopted a uniform prior 
for all parameters except the slope, for which we took instead a uniform prior on the angle.
Assuming a Student-$t$ distribution with 10 degrees of freedom for the scatter,
which is more robust to the presence of outliers, does not change the
results significantly.

While the sample of 14 galaxies is small, it is still five times larger than other samples
of massive ($\log M/M_\odot\gtrsim 11$) cluster galaxies at similar redshift
for which a size has been measured:
IDCS J1426.5+3508 has only
two such quiescent galaxies, for example, and CL J1449+0856 has only one\footnote{We
accounted for differences in
mass definitions across works.}.

To test whether the location of a galaxy in the size--mass plane depends on its stellar age,
points in Figure 8 have been color-coded according to their location in the age--mass plane.
Red and blue symbols show galaxies that are older or younger than the mean age--mass
relation by at least $1\sigma_{intr}$, respectively, while those within $\pm 1\sigma_{intr}$
are black. Note that by comparing ages to the mean age--mass relation we naturally account
for the correlation in these parameters.
There is no evidence that older or younger galaxies have systematically larger or smaller
sizes than the mean at a given mass.
We note that this result is not sensitive to our particular choice of age bins.

\section{Evolutionary trends}

In the previous section we carried out a first comparison between
JKCS\,041 and low--redshift clusters. We found that quenching is fully
in place over 2 orders of magnitude in stellar mass. We also found
that the spatial distribution of red galaxies is similar to that of
nearby clusters and is well described by a $\beta$ profile of normal
core radius. We found no reversal of the color--density (or
cluster-centric distance) relation nor an unusual abundance of
star--forming galaxies in the cluster core, in line with clusters in
the local Universe.  Having used the unique dataset for JKCS\,041 to
carry out unprecedented measurements of the galaxy cluster population
at $z=1.80$ we now construct comparison samples at lower and similar
redshift to further investigate evolutionary trends and cluster--to--cluster 
variation.

\begin{figure}
\centerline{\psfig{figure=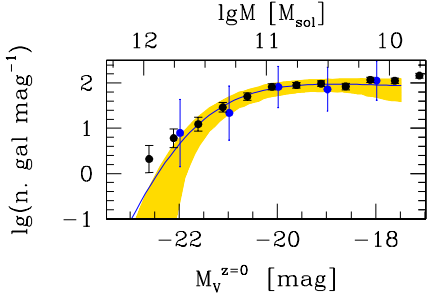,width=6truecm,clip=}}
\caption[h]{Comparison of the luminosity function of red--sequence galaxies for 
JKCS\,041 (blue points with large error bars, 
shading and lines as in previous figure) and Coma 
(black points; from Andreon 2008).  The JKCS\,041 LF has been re-normalized 
to the Coma cluster $\phi^*$ for display purposes only.  Labels on the 
upper x-axis indicate the stellar mass, derived from the $H$--band magnitude 
assuming our standard BC03 SSP settings with $z_f=3$.}
\end{figure}

\subsection{Evolution of the mass function}

We investigated evolutionary trends in the faint--end slope of the 
red--sequence luminosity function (LF) by combining the results for JKCS\,041
with those from 41 clusters taken from the literature for which high--quality 
data are available.  We used the faint-end slope of the
luminosity function as a measure of the relative abundance of faint
galaxies compared with their more luminous counterparts. This is a more
general approach than using other estimators such as the relative
ratio between bright and faint objects (e.g.  De Lucia et al. 2007;
Stott et al. 2008; see also Andreon 2008).

Figure~10 compares the LF of JKCS\,041 with that of the Coma
cluster\footnote{The $H$--band luminosity at $z=1.803$ has been
converted to the {\it present-day} $V$--band luminosity assuming our
standard BC03 setting with $z_f=3$ to account for 10 Gyr of luminosity
evolution. Note that we would have obtained an indistinguishable
result by using the revised version of the Bruzual
\& Charlot (2003) model with a Chabrier IMF (apart from the obvious
common shift in mass of both Coma and JKCS\,041 galaxies).}.
It is remarkable that the two LFs have indistinguishable shapes even though the
look-back time to JKCS\,041 is more than 10 Gyr.
Figure~11 shows the faint--end slope as a function of redshift for a
sample of 42 clusters, JKCS\,041 and 41 clusters taken from the
literature. The sample was chosen to satisfy the following criteria:
the available colors straddle the rest-frame 4000
\AA \ break; common filters are used for cluster and control field
observations; the data are sufficiently deep (i.e. deeper than $-18.2$
mag in $V$ band, as discussed by Andreon 2008, Crawford et al. 2009
and De Propris et al. 2013).  As detailed in Table 1, we used 18 slope
measurements, relative to 28 clusters (two points are average of
several clusters) in the range $0<z<1.3$ (Andreon 2008), 17
measurements at intermediate redshift ($0.2<z<0.55$, De Propris et
al. 2013, 12 additional clusters) and one at $z\sim0.5$ cluster
(Crawford et al. 2009), for a total of 37 measurements of 42 clusters
(including JKCS\,041).  In Appendix~B we list the determinations
excluded because they failed to satisfy these criteria.

\begin{figure}
\centerline{\psfig{figure=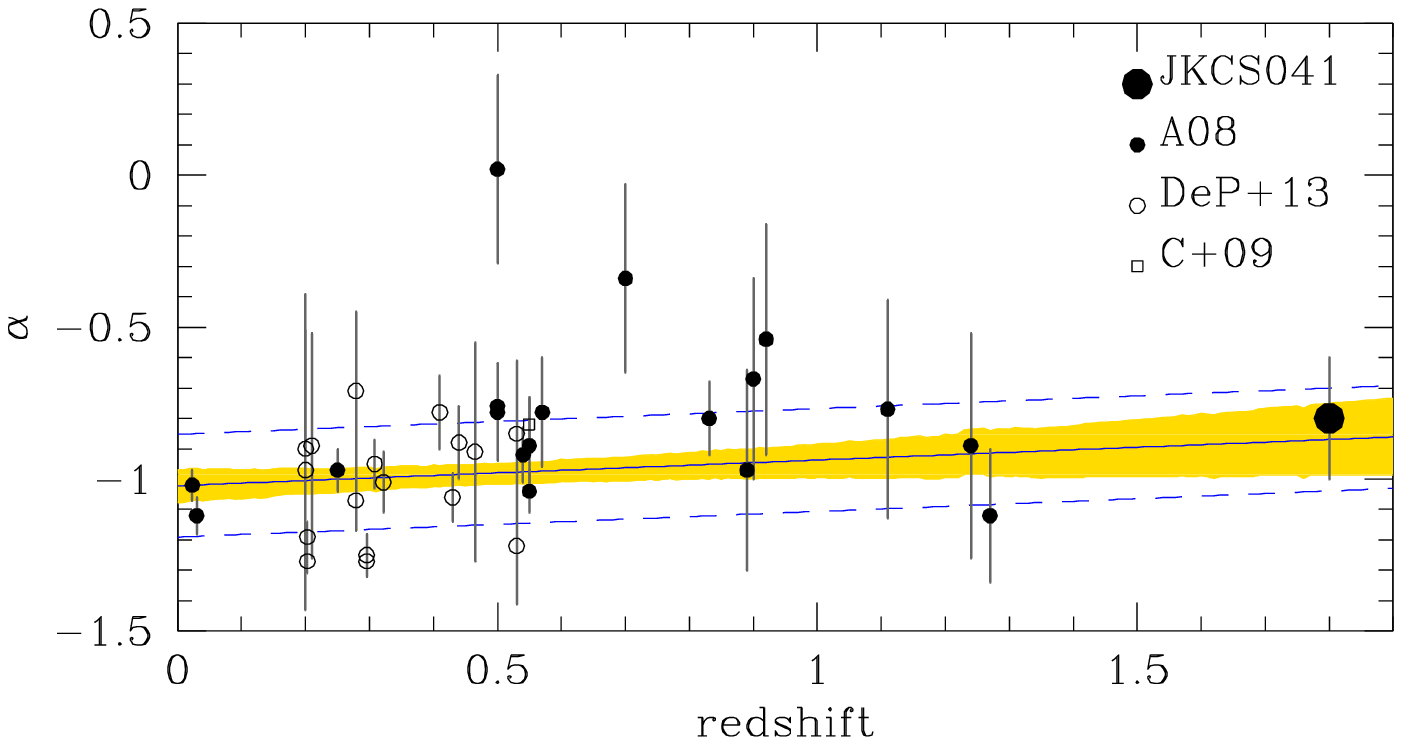,width=9truecm,clip=}}
\centerline{\psfig{figure=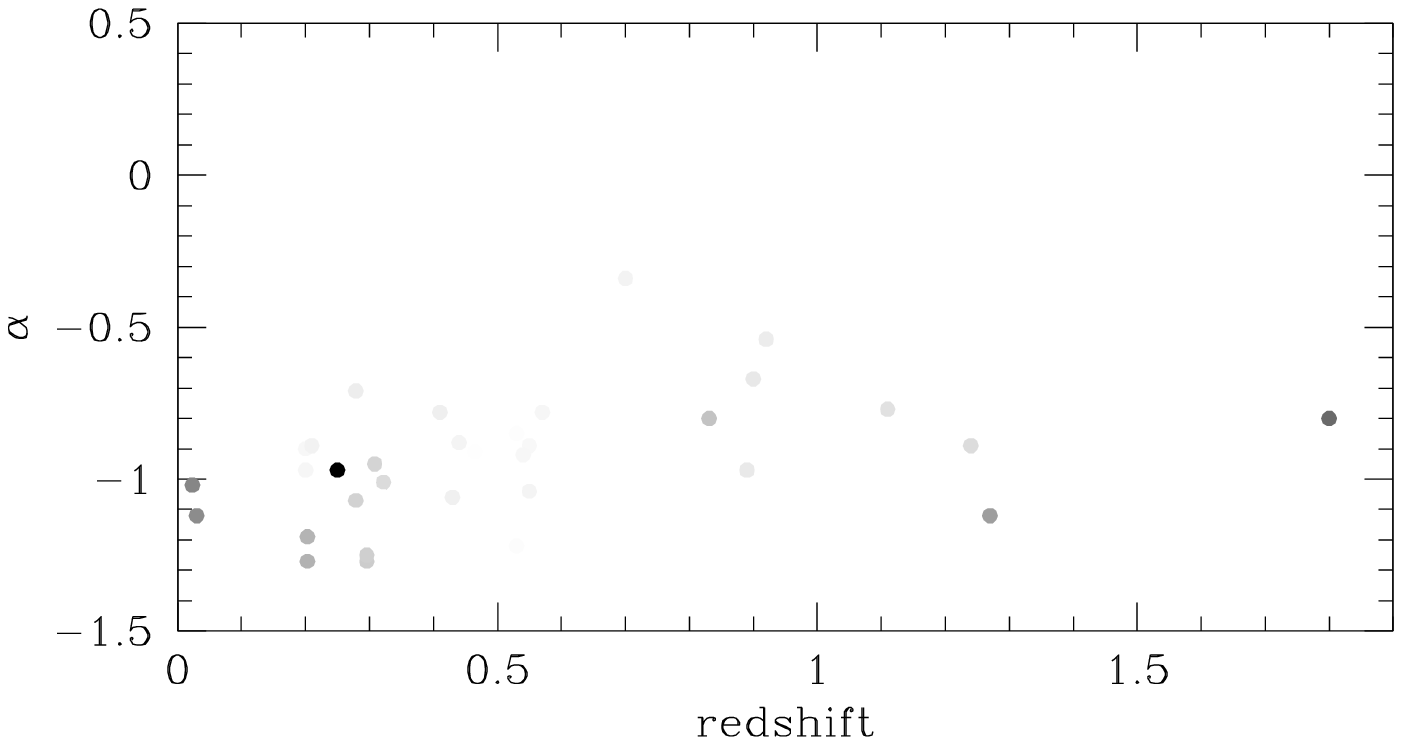,width=9truecm,clip=}}
\caption[h]{Faint--end slope of the luminosity function of red--sequence
galaxies of 42 clusters spanning over 10 Gyr of the age of the
Universe. {\it Top panel:} open dots are from De Propris et
al. (2013), the open square represents the Crawford et al. (2012)
determination, while solid dots are our own work (this work or
Andreon 2008 and references therein).  Some points indicates stacks of
two or ten clusters. The solid line is the mean model, while the
shaded area represent the 68\% uncertainty (highest posterior density
interval) of the mean relation. {\it Bottom panel:} as in the left
panel, with the amount of ink is proportional to the weight given to
each point in the fit.  }
\end{figure}

We fit a linear relation to the data following Andreon
\& Hurn (2013):
\begin{equation}
\alpha(z) = b (z-0.5) + a \ ,
\end{equation}
with an intrinsic Gaussian scatter $\sigma_{intr}$. The parameter $b$
is the evolution of the faint-end slope per unit redshift, while $a$
is the mean faint-end slope at $z=0.5$.  We assumed uniform priors for
the parameters within the physically acceptable ranges, except for the
slope $b$, for which we adopted a uniform prior on the angle (which is
the correct coordinate--independent form; see Andreon \& Hurn 2010).

We find $b=0.08\pm0.09$, $a=-0.98\pm0.03$, and
$\sigma_{scatt}=0.17\pm0.04$ (see Figure~11). The evolutionary term $b$ is
very small and consistent with zero.  Our result is inconsistent with
the evolution suggested by Rudnick et al. (2009) and Bildfell et
al. (2012). The difference arises from the better quality of our data
and the more rigorous selection applied in this study to the select
the clusters, for which the background can be estimated
using the same filters (see Appendix B).

We note that the faint--end slope of the LF has a sizable intrinsic
cluster-to-cluster scatter $\sigma_{scatt}=0.17\pm0.04$. This needs to
be taken into account when analyzing samples of clusters to
avoid spurious results.  Moreover, proper weight
should be given in the fit to each cluster, since the uncertainties
differ widely across the sample. A visual representation of the
importance of each data point is given in the bottom panel of
Figure~11. The weight of each data point is proportional to
that adopted in the fit, which takes into account errors on
measurement, the intrinsic scatter, and the number of clusters per
measurement.  The more informative points are, in strict order of
importance, the stack of 10 clusters at $z=0.25$, the clusters at the
extremes in redshift (JKCS\,041, Coma and Abell 2199), and the stack of
the two Lynx clusters (at $z=1.26$).  This plot qualitatively confirms
the shallow, if any, (and statistically insignificant) evolution of
the faint--end slope over $\Delta z=2$.

We note that our conclusions about the lack of evolution of the faint 
red--sequence galaxies are robust with respect to the methodology. For
example, by using the luminous-over-faint ratio as defined by De Lucia
et al. (2007), we find $L/F= 0.61 \pm 0.17$, consistent with
measurements at lower redshift (Andreon 2008; Capozzi et al. 2010;
Bildfell et al. 2012; Valentinuzzi et al. 2011).
Naturally, several arguments (e.g. Andreon 2008) lead us to expect
evolution for the red sequence, and absence of evidence should not be
interpreted as evidence of absence. Interestingly, the quality of our
data set allows us to set a stringent upper limit to the change of the
faint-end slope $b<0.23$ with 95 \% probability.

Another interesting topic is the evolution of
the bright end of the red sequence.  Most studies of clusters at
$z<1.4$, starting perhaps with Aragon-Salamanca et al. (1993), found it
to be consistent with passive evolution (see also, e.g. Andreon et
al. 2008; De Propris et al. 2013).  Here we take advantage of the
extreme redshift of JKCS\,041 to revisit this measurement in
combination with the high--quality data sample used for the faint--end
slope.

\begin{figure}
\centerline{\psfig{figure=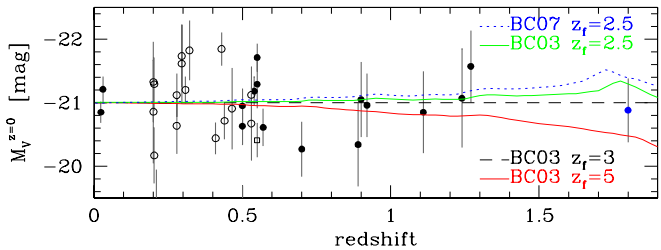,width=9truecm,clip=}}
\caption[h]{Characteristic magnitude of red--sequence galaxies across
10 Gyr. Locii of constant $M^{z=0}_V=-21$ mag for different
formation redshift and Bruzual \& Charlot model version are shown. 
}
\end{figure}

Figure~12 shows the evolution of the characteristic magnitude
$M^{z=0}_{V}$ with redshift\footnote{$M^{z=0}_{V}$ is obtained from
the absolute characteristic magnitude in the band of observation
(close to the $V$-band rest-frame) evolved to $z=0$ and converted into the
$V$ band by adopting a BC03 SSP model with $z_f=3$.}. 
For illustration, a number of model predictions (lines indicating
$M^{z=0}_V=-21$ mag) are shown for different formation redshifts $z_f$
and for the 2007 version of the Bruzual \& Charlot (2003) models. It
is clear that data at $z<1.5$ are insufficient to distinguish between
models.
At the redshift of JKCS\,041, the differences between models are more
significant. Models with $z_f=5$ underpredict the brightness of the
galaxies in JKCS\,041, while models with $z_f=2.5-3$ better
match the data.

With this complete spectroscopic sample showing a fully populated 
red sequence at all masses, we do not confirm the high merger
activity in high--redshift clusters, which was suggested based on the 
statistically estimated paucity
of massive $>L^*$ members on the red sequence (e.g. Mancone et al. 2010).

To summarize, the bright end of the luminosity function of the red
sequence evolves as a passively evolving stellar population formed at
$\sim 2.5<z_f\lesssim 5$. We note that this conclusion does not rely on assuming a
constant faint-end slope and is robust with respect to systematic
uncertainties in stellar population models. More clusters at $z>1.5$
are required to improve this estimate.

\begin{figure}
\centerline{\psfig{figure=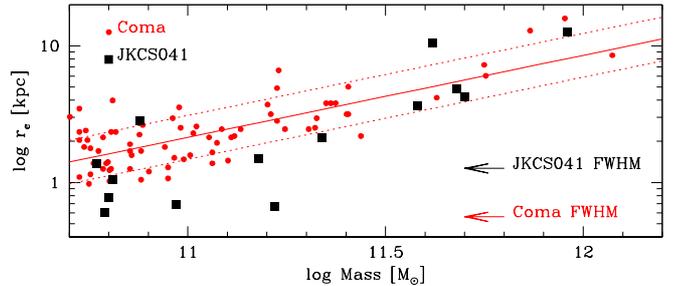,width=9truecm,clip=}}
\caption[h]{Size--mass relation for JKCS\,041 (black squares) and
Coma (red points) galaxies.  The figure also presents the mean fitted
model (solid line) and the one--sigma intrinsic scatter (dotted line)
around the relation for Coma galaxies.  }
\end{figure}

\subsection{Evolution of the galaxy size--mass relation}

As a first approach to understanding evolution of the mass--size
relation, we compared JKCS\,041 with the Coma cluster, chosen because it
has the richest and most controlled set of data: 75 early-type (S0 or
earlier) galaxies more massive than $\log M/M_\odot=10.7$, comprising
a mass--complete sample with high--resolution ($\sim0.6$ kpc)
imaging data (Andreon et al.  1996, 1997). Stellar masses were derived from $R$
magnitudes assuming the standard BC03 setting with $z_f=3$. Effective radiii 
were derived from 
fitting model growth curves\footnote{Note that these are the mean 
observed growth curves of galaxies of different
morphological types, not $r^{1/4}$ growth curves.} from de Vaucouleurs (1977) to
the measured galaxy growth curves (see Michard 1985 for
fitting details). 

The fit of the size--mass relation to Coma galaxies gives
\begin{equation}
\log r_e = 0.33\pm0.02 \ +(0.60\pm0.06)(\log M -11)
\end{equation}
with an intrinsic scatter of $0.16\pm0.01$ dex in $\log r_e$ at a given mass
(see also Figure 13).
This is consistent, within the errors, with the relation found for 
JKCS\,041, as also illustrated by the locus of the  the 68 and 95 \%
probability contours shown in Figure~14. 

The size--mass relation of red--sequence quiescent galaxies in JKCS\,041 is
consistent with the relation for early-type galaxies in clusters at $z\sim0$ within
the uncertainties arising from our sample of 14 systems, which represent
a fivefold increase in the
number of high--redshift galaxies compared with other structures at similar redshifts,
and we place a 95\% confidence upper limit of $0.4$ dex on size growth at fixed stellar mass 
(at $10^{11}$ $M_{\odot}$) between $z=1.8$ and $z=0$, or, equivalently, a $1\sigma$
upper limit of $\sim0.11$ per unit redshift.  

At face value, the best--fit JKCS\,041 mass-size relation indicates a $0.22$ ($=0.33-0.11$)
dex change in the size of cluster galaxies relative to $z=0$, 
in the sense that Coma galaxies are larger than those in JKCS\,041.
In the same redshift range, the mean change in the field from $z=1.8$ to $0$ is 
$0.47$ dex (Newman et al 2012), 
giving a first indication that size evolution might occur earlier in clusters.
The suggested indication is also indirectly implied by
the environmental dependence of the size evolution found in other works
(e.g. Cooper et al. 2011; Lani et al. 2013; Bassett et al. 2013). However,  
this is one of the first works
that presents a direct cluster-to-cluster comparison across redshifts.
A confirmation of this indication would be extremely valuable. 

This comparison clearly relies on the choice of the low--redshift comparison cluster, a few
assumptions, and slightly different methods to derive mass and $r_e$. However,
these do not introduce systematics, as justified in Appendix C.
A more complete analysis of this issue necessitating a comparison with many
clusters along the  $0<z<1.8$ redshift range will be addressed in a future paper.

\begin{figure}
\centerline{\psfig{figure=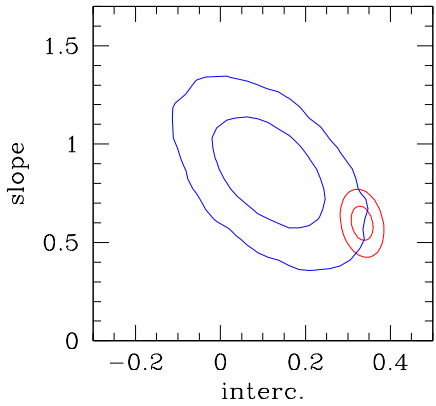,width=6truecm,clip=}}
\caption[h]{68 and 95 \% probability contours for JKCS\,041 (blue)  and Coma 
 (red) size--mass relation.
}
\end{figure}

\begin{table*}
\caption{Cluster redshift, faint--end slope $\alpha$ and characteristic magnitude $M^{z=0}_V$}
\begin{tabular}{l r r c c l}
\hline
Name & $N^{1}$ & z & $\alpha$ & $M^{z=0}_V$  & References \\ 
\hline
JKCS 041 & 1 & 1.80 & $-0.80\pm0.20$ & $-20.88\pm0.50$& This work \\ 
Lynx E+W & 2 & 1.27 & $-1.12\pm0.22$ &  $-21.57\pm0.56$ & A08, This work	     \\  
RDCS J1252-2927 & 1 & 1.24 & $-0.89\pm0.37$ &  $-21.07\pm0.77$  & A08, This work       \\  
RDCS J0910+5422 & 1 & 1.11 & $-0.77\pm0.36$ &  $-20.85\pm0.64$ & A08, This work	     \\  
GHO 1602+4329 & 1 & 0.92 & $-0.54\pm0.38$ &  $-20.96\pm0.50$  & A08, This work	      \\  
GHO 1602+4312 & 1 & 0.90 & $-0.67\pm0.33$ & $-21.05\pm0.59$ & A08, This work	     \\  
1WGA J1226.9+3332 & 1 & 0.89 & $-0.97\pm0.33$ &  $-20.34\pm0.66$ & A08, This work	       \\  
MS1054 & 1 & 0.83 & $-0.80\pm0.12$ &  &  A06, A+08    \\  
MACS J0744.8+3927  & 1 & 0.70 & $-0.34\pm0.31$ &  $-20.27\pm0.43$ & A08, This work       \\  
MACS J2129.4-0741 & 1 & 0.57 & $-0.78\pm0.18$  & $-20.61\pm0.29$ & A08, This work       \\  
MACS J0717.5+3745  & 1 & 0.55 & $-1.04\pm0.07$  & $-21.71\pm0.22$ & A08, This work       \\  
MACS J1423.8+2404 & 1 & 0.55 & $-0.89\pm0.16$  & $-21.29\pm0.38$ & A08, This work	    \\  
CL0016 & 1 & 0.55 & $-0.82\pm0.09$  & $-20.41\pm0.27$ & C+09	     \\  
MACS J1149.5+2223 & 1 & 0.54 & $-0.92\pm0.09$  & $-21.18\pm0.22$  & A08, This work       \\  
J0916+2951 S & 0.5 & 0.53 & $-1.22\pm0.19$   & $-21.12\pm0.44$ & DeP+13     \\  
J0916+2951 W & 0.5 & 0.53 & $-0.85\pm0.24$   & $-20.67\pm0.58$ & DeP+13     \\  
MACS J0911.2+1746 &1 &  0.50 & $-0.78\pm0.16$ & $-20.63\pm0.29$  & A08, This work	 \\  
MACS J2214.9-1359 & 1 & 0.50 & $-0.76\pm0.12$ & $-20.95\pm0.27$  & A08, This work	 \\  
MACS J0257.1-2325 & 1 & 0.50 & $+0.02\pm0.31$  & $-19.42\pm0.26$ & A08, This work 	  \\  
MACS J1621.3+3810 & 1 & 0.46 & $-0.91\pm0.36$ & $-20.91\pm0.64$  & DeP+13	  \\  
MACS J0417.5-1154 & 1 & 0.44 & $-0.88\pm0.12$ & $-20.71\pm0.29$  & DeP+13	  \\  
MACS J0358.8-2955 & 1 & 0.43 & $-1.06\pm0.08$ & $-21.84\pm0.26$  & DeP+13	  \\  
MACS J0553.4-3342 & 1 & 0.41 & $-0.78\pm0.12$ & $-20.44\pm0.25$  & DeP+13	  \\  
A1351 & 1 & 0.32 & $-1.01\pm0.10$  & $-21.82\pm0.47$ & DeP+13	     \\  
A2744 & 1 & 0.31 & $-0.95\pm0.08$  & $-21.20\pm0.21$ & DeP+13	     \\  
Bullet E & 0.5 & 0.30 & $-1.27\pm0.05$ & $-21.73\pm0.49$ & DeP+13     \\  
Bullet W & 0.5 & 0.30 & $-1.25\pm0.07$ & $-21.61\pm0.62$  & DeP+13     \\  
A1758 & 0.5 & 0.28 & $-0.71\pm0.26$  & $-20.64\pm0.44$ & DeP+13	 \\  
A1758 & 0.5 & 0.28 & $-1.07\pm0.10$  & $-21.12\pm0.34$ & DeP+13	  \\  
stack & 10 & 0.25 & $-0.97\pm0.07$  & & A08     \\  
MACS J0547-3904 & 1 & 0.21& $-0.89\pm0.37$ & $-19.18\pm0.77$  & DeP+13      \\  
A2163 & 0.5 & 0.20 & $-1.27\pm0.04$ & $-20.17\pm0.44$  & DeP+13	 \\  
A2163 & 0.5 & 0.20 & $-1.19\pm0.05$ & $-21.29\pm0.42$  & DeP+13	 \\  
A520 & 0.5 & 0.20 & $-0.90\pm0.51$  & $-20.86\pm0.81$ & DeP+13	 \\  
A520 & 0.5 & 0.20 & $-0.97\pm0.46$  & $-21.33\pm0.63$ & DeP+13	 \\  
A2199 & 1 & 0.03 & $-1.12\pm0.06$ & $-21.21\pm0.20$  & A08, This work        \\  
Coma & 1 & 0.02 & $-1.02\pm0.05$  & $-20.85\pm0.17$ & A08, This work	      \\  
\hline                                                      
\end{tabular} \hfill \break
\footnotesize{$^1$ number of clusters per measurement. Values $<1$ indicate
that separate regions are used for the cluster\hfill\break
 \hfill\break}
\end{table*}

\section{Discussion}

\subsection{Comparisons with other high redshift clusters}

Although studies of a single cluster at $z=1.803$ will not allow us to infer the evolutionary properties
of the population of distant cluster galaxies as a whole, the high--quality data available
for JKCS\,041 allow us to perform unique analyses that are currently not 
possible for other clusters
at comparable redshift.

To illustrate the advantages, we compare the red sequence in JKCS\,041 with that
in the $z=1.75$ cluster IDCS J1426.5+3508 (Stanford et al. 2012). The latter data are complete to 
$H\la 22.5$ mag (set by the blue filter), which is unfortunately too bright for assessing the relative
proportion of luminous and faint members. This is the general situation for most
other distant clusters: a direct comparison of the faint end of the red sequence   
is not yet possible because of the lack of similarly deep data in two
bands straddling the 4000 \AA \ break.

Even studies of the luminous end of the cluster red sequence are often hampered 
by confusion regarding cluster membership. For example, IDCS J1426.5+3508 has 
only two spectroscopically confirmed members. Although the magnitude of the brightest 
member is $H=20.4$ mag both in IDCS J1426.5+3508 and JKCS\,041, 
the second--brightest red-sequence (candidate) galaxy in IDCS J1426.5+3508 
is $\ga 0.8$ mag fainter, while JKCS\,041 has three galaxies within this range. Therefore,
quantitative statements on the nature of the red sequence in IDCS J1426.5+3508 must 
await a substantial increase in spectroscopic redshift coverage.
It is also possible to compare with the luminous end of the red sequence in
the CL J1449+0856 structure at $z=1.99$ (Gobat et al. 2009, 2013). However,
there is only one galaxy of similar luminosity (with $H\la22.5$ mag, ``rescaled" to the JKCS\,041 
redshift and accounting for the different $H$ filter used), which implies a significant difference with
JKCS\,041, where we have $\sim12$ cluster galaxies. 

The well-populated luminous red sequence in JKCS\,041 indicates that
the lack of bright red galaxies in candidate high--redshift clusters
(e.g. Mancone et al. 2010, see Andreon 2013)
is not a universal phenomenon. The paucity of star--forming galaxies
with $\log M/M_{\odot}>10.5$ argues against the Brodwin et al. (2013) hypothesis
that $z\gtrsim 1.4$ represents a `firework era' for cluster galaxies (e.g. Hilton et al. 2010;
Fassbender et al. 2011). Brodwin et al. (2013) claimed that 40\% of $\log M/M_{\odot}>10.1$  galaxies
in many clusters are undergoing starbursts with $\gtrsim50$ $M_{\odot}$ yr$^{-1}$) galaxies
deep into the cluster center. This is clearly not the case in JKCS\,041.
However, the quality of the data of JKCS\,041 is unique and unmatched by the
Brodwin et al. (2013) data set: quenched galaxies at the studied 
redshifts might be under--represented as a result of the
insufficient depth of the data\footnote{Quiescient (SSP with $z_f=3$
and Chabrier IMF) galaxies at $z=1.4$ with mass equal to the
mass limit adopted in Brodwin et al. (2013), 
$\log(M/M_\odot)=10.1$, would be between 0.7 and 2 mag fainter than the $5 \sigma$ limit
of their data (Jannuzi \& Dey 1999) in all three bands blueward of
the 4000 \AA \ break.}.
Consequently, the starburst component
may be overestimated. Furthermore, at $r\sim 1.5$ Mpc,
the radial profile of the SFR per unit area seems to flatten at 
$85\pm5$ M$_\odot$ yr$^{-1}$ arcmin$^{-2}$, 
indicating either that the clusters are more extended than this or 
that the background contribution is
underestimated. However,
only three starforming members have been confirmed
in each of the two most distant clusters in Brodwin et al. (2013)
using grism spectroscopy to a depth an
order of magnitude more sensitive in star formation (Zeimann et al.
2013).
On the other end, if the results by Brodwin et al. (2013) 
will be confirmed, this would indicate a cluster-to-cluster 
variance in the effectiveness of
the environmental--dependent quenching.

\subsection{The significance of the JKCS\,041 cluster}

The properties of JKCS\,041 are remarkable in many ways. The cluster mass will likely
grow by a factor of about 3 by $z=0$ (Fakhouri et al. 2010) to emerge locally as a system similar to the  Coma
cluster. Yet, even at $z=1.803$, JKCS\,041 already resembles many nearby clusters: it
has a similar core radius, a tight red sequence with no deficit of faint or bright red
galaxies, few star--forming galaxies in its central region (about $0.7
r_{500}$). Moreover, its age--mass and size-mass relations agree with
those derived in nearby clusters, given the uncertainties set by the
JKCS\,041 sample size. JKCS\,041 contrasts with its adjacent $z=1.8$ field as Coma does with the
local field: from a comparison with field galaxies of the same mass 
at $z\sim1.8$, in Paper I we found that
environmental quenching (i.e. the cluster  environment) is responsible for more than
half the quenched galaxies, quite similar to what is observed at $z\sim0.3-0.45$ and
$z\sim0$ (Andreon et al. 2006). At $z=1.8$, the comparison between quiescent
galaxies in the  field and JKCS\,041 shows that the cluster environment affects the
proportion of quenched galaxies, but not {\it when} they were quenched (Paper I),
an inference similar to  what is found locally. In short, the environment 
determines the space density of each type but not the internal properties of the type, 
which are the same across all environments (Andreon 1996).
These results are surprising when one considers that the last major episode of star formation
was at most 2 Gyr, and more typically 1 Gyr, earlier. In this remarkably short
time JKCS\,041 has already achieved a resemblance to nearby clusters, a similarity
it presumably must maintain during its sizable continued mass growth over the next
10 Gyr. 

We suggest the following scenario to explain the observations: first, cluster growth
involves acquiring additional material with a composition that already resembles that
of the main progenitor and not distributing it randomly in the cluster. Both are well--known features
of the hierarchical paradigm of galaxy formation. Galaxies in a cluster at a given time arise 
from regions of above-average density at all epochs of their history (e.g. Kaiser 1984). 
At low (e.g. Balogh et al. 1999) and intermediate redshift (Andreon et al. 2005) and in
JKCS\,041 (Raichoor \& Andreon 2012a), there is a high fraction of quenched
galaxies well past the virial radius. At low redshift at least, the high--mass end of the 
mass function of galaxies of a given morphology are environment-independent 
(e.g. Bingelli, Sandage \& Tammann 1988; Andreon 1998; Pahre et al. 1998).
Collecting material in a self-similar way helps to maintain a fixed faint-slope of the 
mass function and a fixed characteristic mass, while the luminosity function 
normalization $\phi^*$ and the cluster richness and mass are free to increase. 

The radial dependencies (e.g. the core radius, the lack of star--forming
galaxies in the cluster core) are maintained since, in a hierarchical
picture, the cluster-centric distance distribution of infalling galaxies is not 
uniform. Smith et al. (2012) illustrated this convincingly in the
case of the environmental quenching: galaxies in the cluster core fell inwards
at an earlier times, on average. This preferentially builds up the peripheral
regions, outside the field of view considered here.

After submitting this paper we became aware of Cen (2014), who 
presented state-of-art galaxy formation simulations 
of a region centered on a galaxy
cluster. These simulations reproduce our findings: there is no
lack of faint red galaxies, quenching is found to be more efficient, but not faster 
in dense environments, faint red galaxies are younger
than their massive cousins, 
mass growth occurs mostly in the blue cloud and therefore the characteristic
mass of red sequence galaxies should not strongly evolve with redshift,
correlation between observables of red galaxies, because the age-mass
or mass-size relations are largely inherited and therefore mantained with z.

\section{Summary}

The unique depth of two-color near-infrared HST images (complete above 
an effective stellar mass $\log M/M_\odot=9.8$) 
and spectroscopy (complete to $\log M/M_\odot=10.8$ independently 
of the spectral type) of the $z=1.803$ JKCS\,041 cluster has allowed us
to investigate in detail the cluster mass, the spatial distribution of red galaxies,
the mass/luminosity function, the color--magnitude diagram, and
the age--mass and size--mass relations.

The abundance of high--quality data in JKCS\,041 is apparent by comparing the 
color-magnitude diagram with those of other high--redshift clusters. 
For the first time, it was
possible to eliminate foreground galaxies to $H<22.5$ mag via spectroscopy.
Statistically, it is possible to account for contamination to $H=25$ mag. This
provides the first cluster red sequence at this redshift
over two orders of magnitude in stellar mass. 
Furthermore, the detailed spectroscopy (and 12-band photometry) enables us
to estimate the stellar ages for a complete sample.

Our analyses provided the following results:

\begin{enumerate}

\item {} The JKCS\,041 cluster mass, determined from four different methods
(X-ray temperature, X-ray luminosity, gas mass, and cluster richness), is consistently
$\log M_{200}/M_\odot \gtrsim 14.2$, demonstrating unequivocally that JKCS\,041 
is a massive system that will most likely become a present-day cluster similar to
the Coma cluster.

\item {} The JKCS\,041 red sequence of quiescent members is remarkably well-formed:   
all massive cluster members are red, and there is no deficit of faint red galaxies down
to a stellar mass of $\log M/M_\odot=9.8$. Consequently, quenching has already
occurred over two orders of magnitude in mass. By comparison with field galaxies
at a similar redshift (Paper I), environmental quenching must be responsible for
more than half the red galaxies.

\item{} The red--sequence galaxies in JKCS\,041 are concentrated towards the cluster center
with a spatial distribution well described by a beta profile with core radius 330 kpc. Blue galaxies
represent a minority and avoid the cluster center, as in nearby clusters.  

\item{} The JKCS\,041 red sequence is interpreted as being 
mostly an age sequence with less massive
galaxies significantly younger. 
The age scatter at a given mass is $38\pm9$ \%, that is $370$ Myr 
at $\log M/M_\odot=11$. The star-formation-history weighted age is typically $1.1\pm0.1$
Gyr, corresponding to a formation epoch $z_f=2.6\pm0.1$.

\item{} The size--mass relation of red--sequence quiescent galaxies in JKCS\,041 is
consistent with that observed locally given the uncertainties arising from the
smaller sample size at high redshift. Given that the mean change in the field 
from $z=1.8$ to $0$ is larger ($0.47$ dex vs $0.22$ dex, at face value), 
this is a first indication that size evolution might occur earlier in clusters.

\item {} Complementing JKCS\,041 data with data from 41  clusters at lower
redshift, the {\it shape} of the galaxy mass function of red--sequence galaxies
is unaltered over the last 10 Gyr. We derive stringent upper limits on any change of 
the faint-end slope and characteristic mass of the distribution.

\end{enumerate}

Despite the proximity in time to the quenching era and the likely three-fold increase in mass over
the subsequent 10 Gyr, JKCS\,041 is already remarkably similar to present-day clusters.
We provided a qualitative scenario where this occurs within the context 
of the hierarchical paradigm of galaxy formation.

\begin{acknowledgements}
SA acknowledges Veronica Strazzullo for comments on an early version
of this draft and Stefano Ettori and Fabio Gastaldello for the JKCS\,041 gas mass computation.
We acknowledge HST, program 12927, and CFHT, see full-text acknowledgements at
http://www.stsci.edu/hst/proposing/documents/cp/10\_Proposal\_Implementation12.html
and http://www.cfht.hawaii.edu/Science/CFHLS/cfhtlspublitext.html.
A.R. acknowledges financial contribution from the agreement ASI-INAF
I/009/10/0 and from Osservatorio Astronomico di Brera.
\end{acknowledgements}

{}

\appendix

\section{details of fitting the age--mass relation}

\begin{figure}
\centerline{\psfig{figure=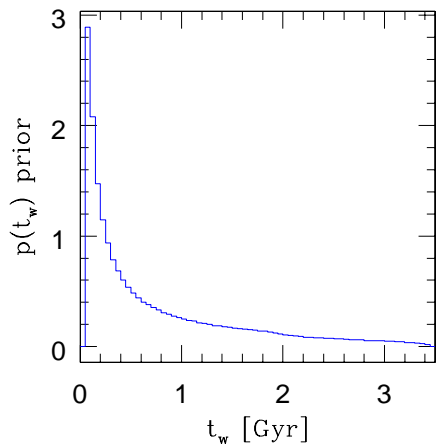,width=6truecm,clip=}}
\caption[h]{Implicit prior on the SFH-weighted age $t_w$ used in Paper I when fitting
galaxy spectra and
photometry, arising from the log-uniform priors on $t_0$ and $\tau$. In the present
paper we prefer a uniform prior on $t_w$ and so have divided out this prior.
}
\end{figure}

Based on the posterior probability distributions for $t_0$ and $\tau$ obtained
in Paper I (section 4) by fitting the grism spectra and 12-band photometry, we computed the
posterior probability distribution of the SFH-weighted age $t_w$ using Eqn.~2.
The posteriors derived in this way assume a prior on $t_w$, shown in Figure~A.1,
that arises from the bounded log-uniform priors used for $t_0$ and $\tau$. 
To base our results for $t_w$ on a uniform prior in that parameter, we
divided the posterior distributions by the prior in Figure~A.1. The
resulting distributions are shown in Figures~7 and 8.

The age--mass posterior probability distribution, $p(t_w,M)$, can be
approximated by a multi-normal distribution with parameters given by the moments
of $p(t_w,M)$, which are outlined in Figure~8 as ellipses. Note that this
approach naturally accounts for the known covariance between the measured mass
and age. The multi-normal approximation is sufficiently accurate for our
purposes, since the scatter in age at a given mass is dominated by intrinsic
galaxy-to-galaxy variation.

We fit the relation between mass and SFH-weighted age $t_w$ with a model that is
linear in the logarithm of the stellar mass and allows for intrinsic scatter,
which is clearly visible when inspecting Figures~7 and 8. Given that ages can
only be positive, we modelled a scatter in $\log t_w$ (scatter in linear units
would allow for unphysical negative ages). To deal with potential
outliers, such as ID 355 (whose young age for its mass was discussed in \S4.4),
we adopted an outlier-resistant model of the intrinsic scatter: a Student-t
distribution of $\log t_w$ with 10 degrees of freedom, which has more extended
tails than a normal distribution. We took a uniform prior on the parameters,
except for the slope, for which we instead used a uniform distribution in the
arctan of the slope (see Andreon \& Hurn 2010 for details). Finally, we compared
the posterior probability distribution using JAGS (Plummer 2008), a standard
MCMC sampler.

\section{Literature measurements of the faint--end slope of the red-sequence
luminosity function}

\begin{figure}
\centerline{\psfig{figure=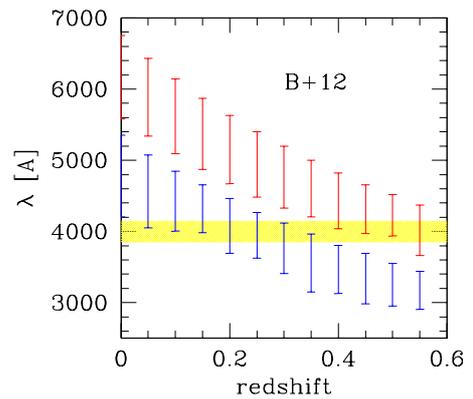,width=6truecm,clip=}}
\caption[h]{Rest-frame $\lambda$ sampling of the adopted filters for the clusters
studied in Bildfell et al. (2012) ($z<0.57$). 
The shaded (yellow) band marks the
4000 \AA \ break.
}
\end{figure}

An accurate measurement of the red-sequence LF requires that the filters used
tightly sample the 4000 \AA \ break, that the same filters are used for both the
cluster and control field observations, and that data are of adequate depth
(Andreon 2008; Crawford et al. 2009; De Propris et al. 2013). Here we comment on
our compilation of literature measurements of the faint--end slope shown in
Figure 11, omitting works already discussed in Andreon
(2008). In particular, we describe the data that we chose to omit because they
do not satisfy the above criteria.

Our compilation of measurements does not included those by Rudnick et al.
(2009), because their data are not deep enough to measure the
faint--end slope (De Propris et al. 2013). Furthermore, their estimation of the
background correction was based on photometry from the Canada--France--Hawaii
telescope in the Mould filters, while their cluster data come from a
different filter set (Bessel filters in use at Paranal). The conclusions by
these authors that the faint end of the red-sequence LF becomes depopulated
toward higher redshifts may instead arise from a combination of these
observational effects. A similar mismatch in the filter set also led us to
exclude two of the clusters studied by De Propris et al.~(2013). One additional
cluster in De Propris et al.~(2013) is already included in the Andreon (2008)
compilation and therfore was excluded to avoid duplication.

Of the five clusters in the Crawford et al.~(2009) sample, only CL0016 met our
depth requirement ($-18.3$~mag) and was included in our analysis. In any
case, data for two of their clusters were superseded by the deeper observations of
the same systems included in the Andreon (2008) sample. We did not include the
faint-end slope of seven clusters at $1<z<1.4$ measured by Mancone et al.~(2012) since their LF is not
corrected for confusion/crawding, which is known to be important for Spitzer data (see
Andreon 2013).

Bildfell et al. (2012) reported $L/F = 0.6$ at $z\approx0.2-0.3$, which agrees
with our measurement in JKCS\,041. However, they claimed a strong evolution of
faint--red sequence galaxies, which is driven by clusters at $z < 0.15$. The
$L/F$ they measured in this low-redshift sample is lower than that found by
Andreon (2008), Valentinuzzi et al. (2011), and Capozzi et al. (2010). As
Figure~B.1 shows, the filters used by Bildfell et al. (2012) do not bracket the
4000 \AA \ break at $z<0.15$. This may be the cause of their discrepant
measurement.

Finally, Lemaux et al. (2012) claimed a deficit of faint red
galaxies in three clusters and groups at $z\sim0.9$ from a qualitative
inspection of their luminosity function. Two of the clusters are in common with
the Andreon (2008) sample, one of which was also studied by Crawford et al.
(2009). With these deeper data, the LF shows a normally populated red sequence,
as quantified by either the faint-end slope or the $L/F$ ratio, although we note
that these two clusters are the most extreme in the present sample.

\section{the size--mass relation at $z=0$ in clusters}

\begin{figure}
\centerline{\psfig{figure=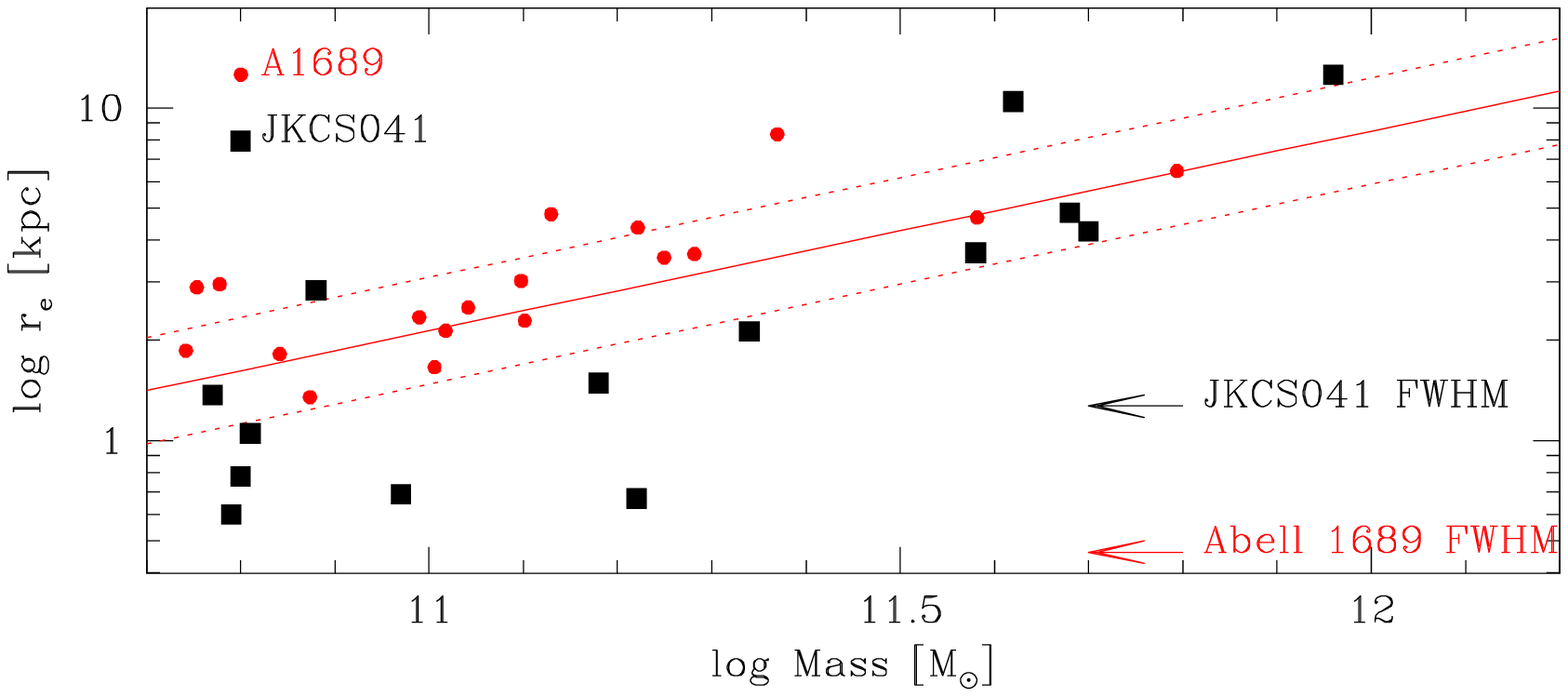,width=9truecm,clip=}}
\centerline{\psfig{figure=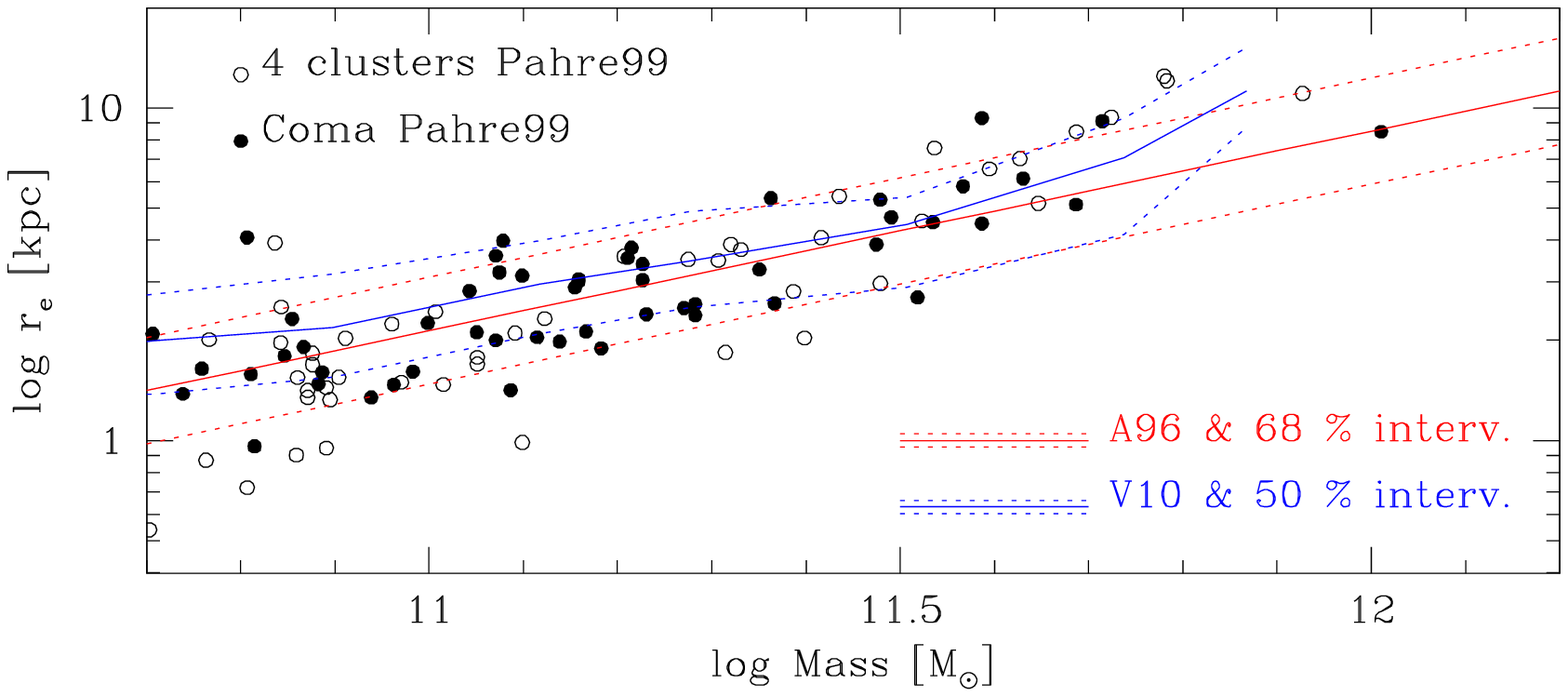,width=9truecm,clip=}}
\caption[h]{Size--mass relation for JKCS\,041 (top panel, black points), A1689
(top panel, red points), and for five
nearby clusters (bottom panel, points) and clusters in Valentinuzzi et al. (2010,
blue lines). Both panels report
the mean fitted model (solid line) and the $1\sigma$
intrinsic scatter (dotted line) for Coma galaxies in Andreon et al. (1996, 1997), our
fiducial
$z = 0$ relation.}
\end{figure}

Although the data used to construct the $z=0$ size--mass relation in \S 4.5 are
of high quality, the methods used to derive stellar mass and $r_e$ at $z = 0$
are not identical to those used in our study of the JKCS\,041 members in
Paper~I. Therefore, it is important to verify that our conclusions are robust to
differences in methodology.

Sizes of the Coma members drawn from Andreon et al.~(1996, 1997) are based on a
curve-of-growth analysis, while the JKCS\,041 sizes were instead derived from
Sersic fits to the 2D light profiles. However, Aguerri et al.~(2004) measured
Sersic-based sizes of Coma members and found these to agree with the Andreon et
al.~(1996, 1997) radii within 10\% for the galaxies in common. Furthermore, the
Augerri et al.~(2004) radii were found to be consistent with later measurements
based on HST images using the Advanced Camera for Surveys (Hoyos et al. 2011),
which cover a smaller region of Coma but have exquisite depth and resolution.
This suggests that the exact method and dataset do not have a strong effect on
the sizes at $z=0$.

In addition to Coma galaxies, for our $z \sim 0$ compilation we also drew upon
measurements in A1689 at $z=0.189$ by Houghton et al. (2012), who assembled a
mass-limited sample of early-type galaxies and used Sersic profile fitting to
derive $r_e$. Furthermore, we included the five nearby clusters studied by Pahre
(1999), including Coma, who derived $r_e$ by fitting seeing-convolved de
Vaucouleurs (1948) profiles to $K$-band imaging. In all cases, we converted from
luminosity to stellar mass using our standard BC03 model with $z_f = 3$. For the
Pahre data we used the $K$-band luminosity, while the $R$-band luminosity was
used for the other clusters.

The fit to the size--mass relation in A1689 gives
\begin{equation}
\log r_e = 0.43\pm0.04 \ +(0.47\pm0.12)(\log M -11)
\end{equation}
with an intrinsic scatter $0.16\pm0.03$ dex in log $r_e$ at a given mass. For the
Pahre (1999) sample we find
\begin{equation}
\log r_e = 0.29\pm0.02 \ +(0.78\pm0.04)(\log M -11)
\end{equation}
with an intrinsic scatter $0.15\pm0.01$ dex in log $r_e$ for the full sample (bottom
panel of Figure~C.1), and
\begin{equation}
\log r_e = 0.26\pm0.03 \ +(0.87\pm0.07)(\log M -11)
\end{equation}
with an intrinsic scatter $0.16\pm0.02$ dex in log $r_e$ when Coma is excluded.
Figure~C.1 demonstrates that these relations are consistent with our fiducial
relation derived for Coma based on the Andreon et al.~(1996, 1997) measurements.
In the lower panel, we also compare this with the size--mass relation from Valentinuzzi
et al.~(2010) and again find reasonable consistency. To summarize, neither the
precise method used to derive mass and $r_e$, nor the particular clusters chosen
for comparison appear to produce strong effects on the size--mass relation of
early-type cluster galaxies at $z = 0$.

We note that the Shen et al. (2003) size--mass relation for early-type field
galaxies in the SDSS, selected using a Sersic index threshold $n > 2.5$, falls
on the upper boundary of the relations plotted in Figure~C.1. Since the present
paper is based on the evolution of cluster galaxies, we preferred to use local
samples that are confined to cluster members. Furthermore, differences in the
morphological distribution of cluster and field galaxies can affect a comparison
with the size--mass relation, as discussed in Paper~I.

\end{document}